\documentclass[acmsmall,screen,nonacm]{acmart}
\setcopyright{none}
\renewcommand\footnotetextcopyrightpermission[1]{}

\newcommand{\ToolName}{\textsc{DAInfer+}}

\newcommand{\Continue}{\textnormal{\textbf{continue}}}

\usepackage[utf8]{inputenc}
\usepackage{fmtcount}
\usepackage{wrapfig}

\newcommand{\cref}{§ \ref}

\usepackage{amsmath,amsfonts}
\usepackage{graphicx}
\usepackage{textcomp}
\usepackage{multirow}
\usepackage{xcolor}
\usepackage[linesnumbered,commentsnumbered,ruled,vlined]{algorithm2e}
\usepackage{algorithmicx}
\usepackage[noend]{algpseudocode}
\usepackage{multicol}

\usepackage[linguistics]{forest}
\usepackage{subcaption}
\usepackage{makecell}
\usepackage[table]{xcolor}
\usepackage{collcell}
\usepackage{pgf}

\newcommand{\ApplySmall}[1]{%
  \pgfmathsetmacro{\percent}{#1*100*0.9}
  \edef\x{\noexpand\cellcolor{green!80!black!\percent!yellow!80!white!}}\x #1%
}
\newcommand{\ApplyLarge}[1]{%
  \if\relax\detokenize{#1}\relax #1\else
  \if\detokenize{#1}- -\else
    \edef\percent{\fpeval{min(100, round((#1 / 100) * 100 * 0.5, 0))}}%
    \edef\x{\noexpand\cellcolor{green!80!black!\percent!yellow!90!white!}}\x #1%
  \fi\fi
}

\newcolumntype{G}{>{\collectcell\ApplySmall}c<{\endcollectcell}}
\newcolumntype{K}{>{\collectcell\ApplyLarge}c<{\endcollectcell}}

\usepackage{booktabs}
\usepackage{caption}
\usepackage{amsthm}
\usepackage{balance}
\usepackage{fmtcount}
\usepackage{tcolorbox}
\usepackage{soul}

\usepackage{listings}
\usepackage{microtype} 

\usepackage{algpseudocode}
\usepackage{pifont}
\usepackage{mathtools}
\usepackage{amsmath}
\usepackage{url}

\usepackage{etoolbox}
\usepackage{mathrsfs}

\usepackage{enumitem}
\usepackage{balance}

\usepackage{bm}
\usepackage{pgf,tikz}

\usepackage{caption}

\usepackage[utf8]{inputenc}
\usepackage{fmtcount}

{\theoremstyle{definition}}
{\theoremstyle{definition}}
{\theoremstyle{definition}}
{\theoremstyle{definition}\newtheorem{definition}{Definition}}
{\theoremstyle{definition}\newtheorem{example}{Example}}

\newif\ifshowrevisioncomments
\showrevisioncommentsfalse
\ifshowrevisioncomments
\newcommand{\revision}[1]{\textcolor{blue}{{#1}}}
\newcommand{\deletion}[1]{\textcolor{red}{{\st{#1}}}}
\else
\newcommand{\revision}[1]{#1}
\newcommand{\deletion}[1]{}
\fi

\newif\ifshowcomments
\showcommentstrue
\ifshowcomments
\newcommand{\maryam}[1]{\mytodogreen{[maryam: #1]}}
\newcommand{\zhouan}[1]{\mytodoblue{[zhouan: #1]}}
\else
\newcommand{\maryam}[1]{}
\newcommand{\zhouan}[1]{}
\fi

\newcommand{\mytodoblue}[1]{\textcolor{blue}{\ding{46}~{\sf}~#1}}

\newcommand{\mytodogreen}[1]{\textcolor{cyan}{\ding{46}~{\sf}~#1}}

\newcommand{\mybox}[1]{
	\begin{tcolorbox}[
		boxsep=-0.5pt,
		standard jigsaw,
		boxrule=0.6pt,
		opacityback=0,
		sharp corners]
		#1
	\end{tcolorbox}
}

\setcopyright{rightsretained}
\acmJournal{TOSEM}

\begin{document}

\title[]{\ToolName: Neurosymbolic Inference of API Specifications from Documentation via Embedding Models }



\author{Maryam Masoudian}
\orcid{0000-0003-0617-5322}
\affiliation{%
	\institution{The Hong Kong University of Science and Technology}
	\city{Hong Kong}
	\country{China}}
\email{mamt@cse.ust.hk}

\author{Anshunkang Zhou}
\orcid{0000-0001-8719-1070}
\affiliation{%
	\institution{The Hong Kong University of Science and Technology}
	\city{Hong Kong}
	\country{China}}
\email{azhouah@ust.hk}

\author{Chengpeng Wang}
\orcid{0000-0003-0617-5322}
\affiliation{%
	\institution{The Hong Kong University of Science and Technology}
	\city{Hong Kong}
	\country{China}}
\email{cwangch@connect.ust.hk}

\author{Charles Zhang}
\orcid{0000-0001-6417-1034}
\affiliation{%
	\institution{The Hong Kong University of Science and Technology}
	\city{Hong Kong}
	\country{China}}
\email{charlesz@cse.ust.hk}

\begin{abstract}
Modern software systems heavily rely on various libraries, which require understanding the API semantics in static analysis. However, summarizing API semantics remains challenging due to complex implementations or unavailable library code.
This paper presents \ToolName, a novel approach for inferring API  specifications from library documentation. 
We employ Natural Language Processing (NLP) to interpret informal semantic information provided by the documentation, which enables us to reduce the specification inference to an optimization problem.
\textcolor{black}{Specifically, we investigate the effectiveness of sentence embedding models and Large Language Models (LLMs) in deriving memory operation abstractions from API descriptions. These abstractions are used to retrieve data-flow and aliasing relations to generate comprehensive API specifications.}
To solve the optimization problem efficiently, we propose neurosymbolic optimization, yielding precise data-flow and aliasing specifications. 
Our evaluation of popular Java libraries shows that zero-shot sentence embedding models outperform few-shot prompted LLMs in robustness, capturing fine-grained semantic nuances more effectively. While our initial attempts using two-stage LLM prompting yielded promising results, we found that the embedding-based approach proved superior. Specifically, these models achieve over 82\% recall and 85\% precision for data-flow inference and 88\% recall and 79\% precision for alias relations, all within seconds.
These results demonstrate the practical value of \ToolName\ in library-aware static analysis.
\end{abstract}



\maketitle
\section{Introduction}

In modern programming languages, programmers often develop their applications based on various libraries, which provide fundamental building blocks for client-side implementation.
Undoubtedly, the behaviors of library APIs directly affect the functionality of the application code.
As targeted by existing studies~\cite{Bastani0AL18,EberhardtSRV19},
several library APIs are essentially generalized store and load operations, 
forming aliasing relations through store-load matches.
For example, the APIs \textsf{HashMap.put} and \textsf{HashMap.get} conduct the store and load operations, respectively.
When they are invoked upon the same \textsf{HashMap} object with the same first parameters successively,
the return value of \textsf{HashMap.get} can be aliased with the second parameter of \textsf{HashMap.put}.
To identify value flows in the application code, 
a static analyzer should be aware of such API aliasing specifications,
which play critical roles for pointer analysis and other downstream clients.
According to our investigation, many existing static analysis techniques rely on manually specified library API aliasing specifications~\cite{Arzt14FlowDroid, Pratik20SemanticModel, Yannis20SoundModolo}. 
However, the emergence of third-party libraries introduces a large number of APIs, making this laborious effort unacceptable in practice.

This work initially targets the API aliasing specification inference problem to support library-aware alias analysis. 
Existing approaches infer API aliasing specifications from three perspectives.
The first line analyzes the source code statically~\cite{ArztB16, rountev2008ide}.
Although it can derive the function summaries as the API aliasing specifications,
the solution suffers from the scalability problem due to deep call chains~\cite{TomanG17}.
More importantly, the implementation of several library APIs can depend on native code, such as \textsf{System.arraycopy} in the implementation of \textsf{java.util.Vector},
which makes static analysis intractable~\cite{Bastani0AL18}.
The second line of the techniques constructs unit tests via active learning to trigger the execution of library APIs, so as to infer aliasing relations in the runtime~\cite{Bastani0AL18}.
Compared to static analysis-based inference techniques,
they are more applicable when the source code of the library is unavailable.
However, it can be infeasible to generate unit tests to trigger the target library APIs due to the difficulties of constructing the parameters with complex data structures and executing APIs on specific devices or environments. 
Third, several researchers learn the aliasing specifications from applications using libraries~\cite{EberhardtSRV19},
which does not require the source code of the libraries or the execution of the programs.
Unfortunately, their approach only discovers the API specifications used in the applications, 
finally causing the low recall in the inference.

\begin{figure*}[t]
	\centering
	\includegraphics[width=\linewidth]{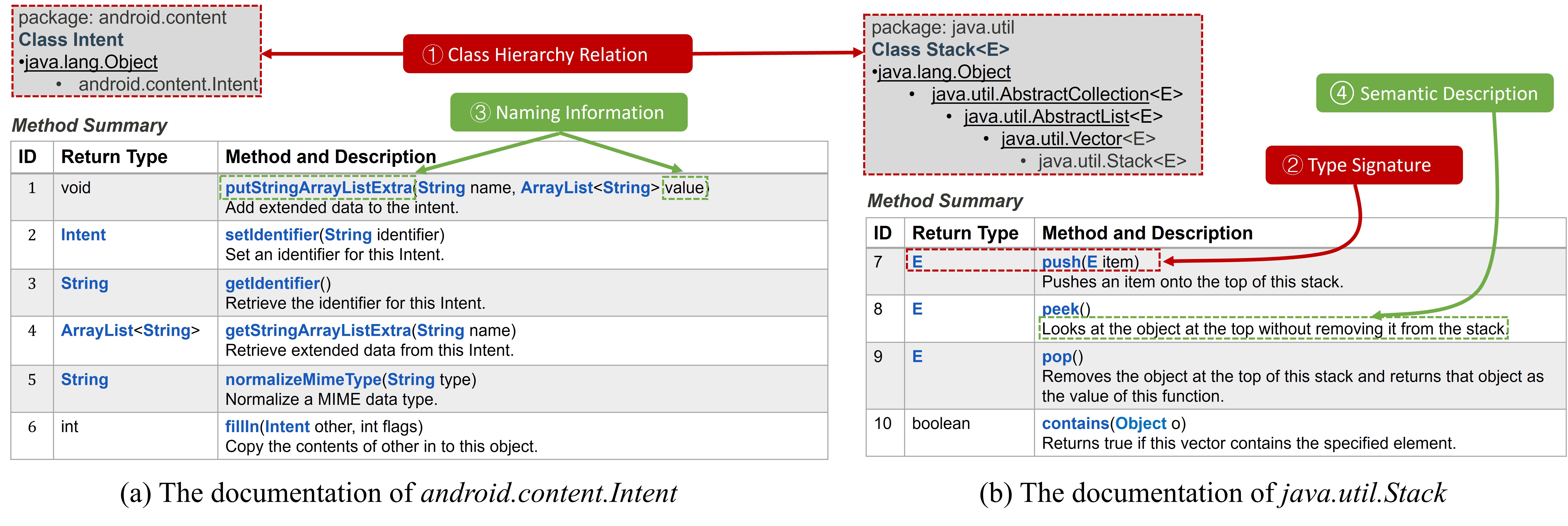}
	\vspace{-6mm}
	\caption{Examples of library documentation. We use $m_i$ to denote the API with the ID $i$ in the paper.}
	\label{fig:javadoc}
	\vspace{-6mm}
\end{figure*}

This paper presents a new perspective on inferring API aliasing specifications. Unlike existing studies, we utilize another important library artifact: the documentation, to analyze the semantics of library APIs.
As shown in Figure~\ref{fig:javadoc}, library documentation contains formal semantic properties,  e.g., class hierarchy relation and type signatures,
and informal semantic information, e.g., semantic descriptions and naming information.
Although the library documentation demonstrates the library API semantics in detail,
it is far from trivial to derive API aliasing specifications from it.
First, effectively understanding the informal semantic information is quite difficult.
Even if we apply the recent advances in the large language models \revision{(LLMs)},
e.g., feeding the documentation of \textsf{android.content.Intent} to \textsc{ChatGPT},
we can only obtain nine API aliasing specifications, all of which are incorrect.
Second, library documentation \revision{can be quite lengthy}, \revision{which may introduce significant overhead.}
For example, feeding the lengthy documentation to \textsc{ChatGPT} not only demands much time but also introduces a high financial cost due to enormous token consumption.

Beyond the cost, we observe a fundamental reliability gap. While LLMs show promise in software engineering tasks such as programming \cite{codegeneration,codingwithllm1,feng-etal-2020-codebert,codesearchnet,APIGen,clear,methodNameRecommendation1}, program analysis \cite{repoaudit2025,LLMDFA10.5555/3737916.3742097,usenixCodeAnalysisLLM,binarytaint10.1145/3711816}, and program repair \cite{programrepair1,programrepair2}, they are prone to semantic over-engineering and hallucinations that compromise their reliability. For instance, an LLM may incorrectly attribute a memory \textsf{write} operation to the simple API method \textsf{Stack.contains} despite the method's signature and documentation clearly indicating it only reads from the stack. Advanced prompt engineering techniques do not consistently mitigate these errors; indeed, the two-staged prompting approach in our previous work~\cite{DAInfer} mistakenly attributes a memory deletion to this method. Furthermore, LLMs struggle to isolate distinct operations within compound sentences in an API method's semantic description, such as the ``\textsf{removes the object... and returns...}'' phrasing of \textsf{Stack.pop} shown in Figure~\ref{fig:javadoc}. For instance, an LLM may focus exclusively on the removal of an item from the stack while failing to infer the data-flow link to the return value. In the context of taint analysis, this oversight creates a broken propagation chain.

These failures in distinguishing conjoined operations, coupled with high computational overhead, necessitate a shift toward more robust methodologies. We argue that sentence embedding models~\cite{sentence-bert,wang2024multilinguale5textembeddings} provide a superior foundation for this task. By focusing on semantic similarity within atomic, single-intent sentences, these models can robustly identify memory operations while maintaining computational efficiency and structural precision.

To effectively achieve the inference with high efficiency, we propose our inference algorithm named \ToolName\, which originates from three key insights:

\begin{itemize}[leftmargin=*]
\item The class hierarchy determines the available APIs of a given class, while type signatures enable us to over-approximate aliasing facts based on the types of API parameters and returns.
If two values can not be aliased,
we do not need to analyze the naming information and semantic descriptions,
which decreases the overhead by avoiding applying NLP models.

\item The named entities in the names \revision{of APIs and parameters} indicate the high-level semantics\deletion{ of the parameters and return values} \revision{and} narrow down aliasing relations\revision{  between the parameters and return values}.
In Figure~\ref{fig:javadoc}(a), the named entities in \revision{\textsf{getIdentifier}} and the parameter name of \textsf{Intent.setIdentifier} are the same, indicating that the return value of \textsf{Intent.getIdentifier} can be aliased with the parameter of \textsf{Intent.setIdentifier}.

\item Semantic descriptions reveal memory operations through specific verbs, supporting the identification of store-load matches that may introduce the derivation of aliasing facts. In Figure~\ref{fig:javadoc}(b), verbs such as \textsf{push} and \textsf{look} indicate that \textsf{Stack.push} and \textsf{Stack.peek} perform insertion and read operations, respectively. 
\end{itemize}

Based on our insights, we propose \ToolName, an algorithm to infer API specifications by finding data-flow and aliasing relations.
Technically, we introduce a graph representation to over-approximate the aliasing relations between parameters and return values based on type information.
To interpret informal semantic information, 
we use NLP models to abstract memory operation kinds and high-level semantics of API parameters/return values, respectively.
We formulate the task as mapping from semantic information from API documentations to formal memory behaviors that govern data-flow and aliasing relations.
We leverage a tagging model to infer the alias relations between return values and parameters of APIs.
Then, we reduce the specification inference problem to an optimization problem that enforces the aliasing pairs between API parameters as many as possible for precise semantic abstraction.
Particularly, the optimization problem poses constraints on the results of the two NLP models.
To solve the problem efficiently, we propose the neurosymbolic optimization algorithm, which interacts with the two NLP models in a demand-driven manner, achieving low resource cost in the inference. 

To accurately infer memory operations for each API method, we previously proposed a staged prompting technique using generative LLMs~\cite{DAInfer}. In this paper, we introduce a more robust solution utilizing zero-shot sentence embedding models for semantic mapping. While the prompting approach tasks the LLM with simulating a developer to categorize API behaviors based on documentation, the embedding approach calculates the cosine similarity between API descriptions and standardized memory operation definitions (e.g., \textsf{read}, \textsf{write}, \textsf{insert}, \textsf{delete}). Our evaluation (Section~\ref{def:moa}) demonstrates that embedding models achieve significantly higher recall and precision than LLMs using either zero-shot or few-shot prompting. By reducing the inference task to a semantic similarity comparison, we eliminate the hallucination risks associated with generative models while achieving greater efficiency.

We implement our approach \ToolName\ and evaluate it upon Java classes in several popular When the two-stage prompting is selected to retrieve memory operation abstractions,\ToolName\ achieves the alias specification inference with a precision of 79.78\% and a recall of 82.29\%,
consuming 5.35 seconds per class on average.
Additionally, \ToolName\ promotes the alias analysis by discovering 80.05\% more aliasing facts for the API return values and enables the taint analysis to discover 85 more taint flows in the experimental subjects.
Furthermore, we assess \ToolName\ capability in retrieving data-flow specifications using embedding models vs. LLMs. Our results demonstrate its recall and precision of embedding models are more than 82\% and 85\%, respectively. In contrast, programming-trained LLMs could only reach up to 75\% recall, albeit with a higher precision of 94\%. This suggests that while specialized LLMs can be more precise, embedding models may provide a more comprehensive retrieval of specifications, which is crucial for effective static analysis. Furthermore, when employing embedding models,  \ToolName\ achieves 88\% recall and 79\% precision in inferring alias specifications while maintaining high efficiency by retrieving results in only a few seconds.
The main contributions of this work are:
\begin{itemize}[leftmargin=*]
\item We propose a novel neurosymbolic optimization technique to solve the API specification inference problem efficiently.

\item We introduce a comprehensive pipeline to perform memory operation inference, which serves as a foundational layer for subsequent alias inference.

\item We introduce a new embedding-driven paradigm for inferring API specifications, replacing generative LLM prompting with a deterministic, embedding-based retrieval mechanism that leverages latent vector comparisons between API descriptions and memory operation abstractions.

\item We conduct a comprehensive comparative analysis between general-purpose LLMs and state-of-the-art programming-specific models, demonstrating that our embedding-based approach achieves superior efficiency and precision in API specification inference. 

\item We extensively evaluate our approach over real-world libraries to demonstrate its improved accuracy and efficiency compared to existing techniques, and to quantify its impact on client analyses.
\end{itemize}

\section{Background and Overview}
\label{sec:overview}

In this section, we introduce the background of API \textcolor{black}{data-flow and} aliasing specification inference and outline our key ideas to \textcolor{black}{infer data-flow and aliasing relations of API methods from their documentations}.

\subsection{Library-Aware \textcolor{black}{Data-Flow and} Alias Analysis}
\label{subsec:background}
Modern software systems heavily depend on various libraries. 
A recent study found that a Java project can include an average of 48 libraries transitively~\cite{WangWLWWYYZC18}. 
This prevalence of library usage stimulates the demand for modeling API semantics in fundamental static analyses, such as \textcolor{black}{data-flow and } alias analysis. 
However, the deep call chains and unavailable source code (e.g., native functions) complicate the scalability and applicability of static analysis.
Many static analyzers use specifications to abstract the library API semantics to achieve library-aware analysis.

A data-flow specification for an API $m$ presents the flows of data from its parameters to its body (acting as a data sink) or from its body to the return value (acting as a data source). This specification primarily represents the memory operations that a method performs upon its execution, such as \textsf{read}, \textsf{write}, \textsf{insert}, and \textsf{delete}.

\begin{example}
     \textcolor{black}{Figure~\ref{fig:javadoc}(a) indicates that the first parameter of \textsf{Intent.putStringArrayListExtra} is used to insert a new String list into an ``\textsf{Intent}'' object, whereas \textsf{Intent.getStringArrayListExtra} retrieves the list from the same object if invoked successively. In a data-flow context, the former method performs an insertion memory operation on the internal state of the ``\textsf{Intent}'' object, while the latter performs a read operation on the same object; this establishes a potential taint path through the ``\textsf{Intent}'' container.
}
\end{example}

By identifying the specific memory operations within these individual specifications, one can derive higher-order relationships between multiple APIs.
Specifically,
the API aliasing specification for an API pair $(m_1, m_2)$ is established by matching their respective memory behaviors:
When $m_1$ and $m_2$ conduct the store and load operations, respectively, 
the return value of $m_2$ may be aliased with the parameter of $m_1$
if $m_2$ is invoked after $m_1$ upon the same object.
Based on the specification, a static analyzer can model the library API semantics without explicitly analyzing the implementation of $m_1$ and $m_2$,
ultimately promoting the scalability and applicability of the overall analysis.

\begin{example}
    Figure~\ref{fig:javadoc}(a) indicates that when the first parameters of \textsf{Intent.putStringArrayListExtra} and \\ \textsf{Intent.getStringArrayListExtra} are aliased, the return value of the latter can be aliased with the second parameter of the former if they are invoked successively upon the same ``\textsf{Intent}'' object.
\end{example}

\subsection{Different Perspectives of Inferring API Specifications}

With the increasing number of third-party libraries,
manually specifying the API specifications demands incredibly laborious effort~\cite{Arzt14FlowDroid, Yannis20SoundModolo, Pratik20SemanticModel}.
To mitigate this problem, previous studies infer data-flow and aliasing API specifications from different artifacts, including library implementation~\cite{ArztB16}, application code using libraries~\cite{EberhardtSRV19,modelgen}, and tests synthesized via active learning~\cite{Bastani0AL18} or coverage-guided fuzzing \cite{spectre}.
However, their solutions can be hindered by three main drawbacks. First, analyzing the library implementation suffers from the scalability issue due to complex program structures, such as deep call chains, and can even become inapplicable due to the unavailability of the implementation or the presence of native code.
Second, inferring the specifications from the application code using libraries may fail to achieve high recall when specific APIs are not utilized in the application code.
Third, deriving the data-flow or aliasing facts from dynamic execution suffers from the inapplicability issue when it is infeasible to construct executable tests in specific devices or environments.

To fill the research gap,
our work proposes another perspective to infer the API specifications.
We realize that there is another essential library artifact, i.e., library documentation,
demonstrating the library API semantics in a semi-formal structure.
As shown in Figure~\ref{fig:javadoc},
the formal semantic properties,
including class hierarchy relation and type signatures,
are explicitly provided.
Meanwhile, 
the naming information, e.g., the parameter names and API names,
shows the intent of API parameters and return values, while semantic descriptions demonstrate the functionalities of the APIs informally.
These ingredients permit us to understand how the library APIs manipulate the memory. 
\textcolor{black}{Specifically, this enables the inference of data-flows to and from the heap upon method invocation, which in turn facilitates the identification of aliasing relations between parameters and return values. }
More importantly, the documentation is often available for analysis,
as the developers tend to refer to it during the development.
Hence, inferring the API data-flow and aliasing specifications from documentation would exhibit better applicability than the existing techniques.

\subsection{Overview of \ToolName}

Although the documentation guides the developers in understanding the API semantics,
there exists a gap between the API knowledge and API \textcolor{black}{data-flow and} aliasing specifications.
Concretely, we need to understand how the API parameters are stored and how the API return values are loaded.
However, achieving this is quite complicated in front of informal semantic information.
Even if we leverage the new advances in the \revision{LLMs}\deletion{large language models},
they cannot understand how the APIs manipulate memory and eventually fail to identify the aliasing relations between API parameters and return values.
\revision{With lengthy documentation, frequent interactions with LLMs can incur significant time and token costs.}
\deletion{Also, interacting with the LLMs via online requests can bring quite high overhead and 
consume a large number of tokens in the presence of long documentation.}

To address the challenges, we propose a novel inference algorithm named \ToolName, 
which effectively understands the API semantics and efficiently infers the API specification from library documentation.
Our key idea originates from three critical observations on the \textcolor{black}{data-flow and} aliasing relations between the parameters and return values of the library APIs as follows.
\begin{itemize}[leftmargin=*]
\item The parameters and return values should be type-consistent if they are aliased. Specifically, their types should be the same, or one of them is the subtype/super-type of the other. 
Such facts can be easily obtained from the class hierarchy relation and type signatures in the documentation. 
In Figure~\ref{fig:javadoc}, for example, we can obtain the potential aliasing relation between the return value of \textsf{Intent.getIdentifier} and the parameter of \textsf{Intent.setIdentifier}, while the second parameter of \textsf{Intent.putStringArrayListExtra} can not be aliased with the return value of \textsf{Intent.getIdentifier}.   

\item  If the return values and parameters of two APIs are aliased, the named entities in their names tend to be the same, indicating the same high-level semantics.
For example, the APIs \textsf{Intent.setIdentifier} and \textsf{Intent.getIdentifier} in Figure~\ref{fig:javadoc}(a) share the same named entity ``\textsf{identifier}'', indicating that they manipulate the same inner field.
For general-purpose data structures, such as \textsf{java.util.Stack} in Figure~\ref{fig:javadoc}(b),
the API names of \textsf{Stack.peek} and \textsf{Stack.pop} do not have any named entities,
indicating that their return values can be aliased with other parameters with consistent types.

\item If a library API stores its parameters or loads the inner field as the return value, the verbs in its semantic description can reflect the memory operation kind intuitively.
For example, the verbs ``set\text{''} and ``insert\text{''} are commonly used for the APIs storing \revision{their}\deletion{its} parameters, while the verbs ``get\text{''} and ``return\text{''} are prevalent in the semantic descriptions of the APIs loading inner fields.

\end{itemize}

\begin{figure*}
	\centering
		\vspace{-2mm}
	\includegraphics[width=0.9\textwidth]{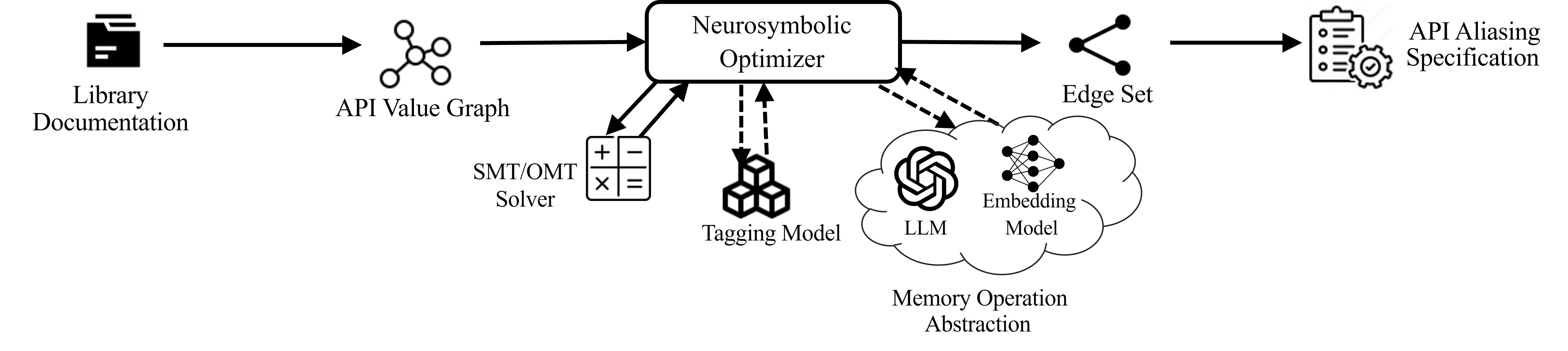}
	\vspace{-4mm}
	\caption{Workflow of \ToolName}
	\vspace{-5mm}
	\label{fig:workflow}
\end{figure*}

Based on the observations, we realize that we can leverage type information to over-approximate aliasing relations and utilize named entities, verbs, and \textcolor{black}{descriptive simple sentences} to understand the high-level semantic meanings of the APIs \textcolor{black}{and their data-flow facts}.
For any store-load API pair, we can finalize an API aliasing specification
as long as we discover the parameters and return values with the same semantic meanings and consistent types.
According to these insights, we design our inference algorithm \ToolName,
of which the workflow is shown in Figure~\ref{fig:workflow}.
Our key technical design consists of three components.

\begin{itemize}[leftmargin=*]
\item We introduce a new graph representation, namely the \emph{API value graph},
to approximate aliasing relations.
After converting a library documentation to a normalized documentation model, we encode the potential aliasing relations in the API value graph.

\item We reduce the \revision{inference problem} to an optimization problem upon the API value graph, where we aim to discover as many aliasing facts among parameters and return values as possible.
Particularly, we leverage NLP models to extract the named entities and interpret the semantic descriptions to \textcolor{black}{infer memory operation abstractions, respectively. These abstractions represent the underlying data-flow facts for each API method, serving as the foundation for our optimization-based inference.}

\item We instantiate the optimization problem and propose an efficient neurosymbolic optimization algorithm to solve the problem,
of which the solution induces the API aliasing specifications.
Our neurosymbolic optimization algorithm interacts with the tagging model and \textcolor{black}{the memory operation abstraction module} in a demand-driven manner,
significantly improving the efficiency of our algorithm.
\end{itemize}

Benefiting from our insights, our inference algorithm \ToolName\ simultaneously achieves high precision, recall, and efficiency.
The high availability of library documentation also promotes the applicability of our approach in real-world scenarios.
In the following sections,
we will formulate our problem (\cref{sec:pf}) and provide our technical design (\cref{sec:abstraction} and~\cref{sec:solving}) in detail.
\section{Problem Formulation}
\label{sec:pf}

This section first formulates the documentation model (\cref{subsec:syntax})
and then defines the API aliasing specification (\cref{subsec:depmodel}).
Lastly, we provide the formal statement of the API aliasing specification inference problem and highlight the technical challenges (\cref{subsec:ps}).

\subsection{Documentation Model}
\label{subsec:syntax}

\begin{definition}(Documentation Model)
\label{def:doc-model}
Given a library,
its documentation model is 
$\mathbf{L}:=(\mathbf{H}, \mathbf{T}, \mathbf{N}, \mathbf{D})$:
\begin{itemize}[leftmargin=*]
	\item Class hierarchy model $\mathbf{H}$ maps a class $c$ to a set of classes, which are the superclasses of $c$.
	\item Type signature model $\mathbf{T}$ maps $(c, m, i)$ to a type, where $m$ is an API of the class $c$ and $i$ is the index of the parameter.
	Without ambiguity, we regard the index of the return value as -1.
	\item Naming model $\mathbf{N}$ maps $(c, m, i)$ to a string indicating the parameter name or API name, where $m$ is an API of the class $c$ and $i$ is the index of the parameter.
	Without ambiguity, $\mathbf{N}(c, m, -1)$ indicates the name of the API $m$ of the class $c$.
	\item Description model $\mathbf{D}$ maps $(c, m)$ to a string indicating the API semantic description.
\end{itemize}
\end{definition}

\begin{example}
According to the documentation of the class \textsf{Intent} in Figure~\ref{fig:javadoc},
we have
$$\mathbf{H}(\textsf{Intent}) = \{  \textsf{Object}   \}, \ \ \mathbf{T}(\textsf{Intent}, m_1, -1) = \textsf{void}, \ \ \mathbf{T}(\textsf{Intent}, m_1, 1) = \textsf{ArrayList<String>}$$
$$\mathbf{N}(\textsf{Intent}, m_1, 0) = \textsf{name}, \ \ \mathbf{N}(\textsf{Intent}, m_1, 1) = \textsf{value}, \ \ \mathbf{N}(\textsf{Intent}, m_1, -1) = \textsf{putStringArrayListExtra}$$
$\mathbf{D}(\textsf{Intent}, m_1)$ is ``\textsf{Add extracted data to the intent}\text{''}.
Here, $m_1$ is the API \textsf{Intent.putStringArrayListExtra}.
Due to space limits, we do not discuss other APIs in detail.
\end{example}

\revision{Based on documentation}\deletion{Notably}, 
we can collect all the APIs offered by a specific class and its superclasses,
forming the universe of available APIs when using the class.
The naming information and API semantic descriptions are informal specifications,
guiding the developers to use proper APIs in their programming contexts.
Based on the documentation model,
not only do developers achieve their program logic conveniently, but also analyzers can understand the behavior of each API.

\subsection{API Aliasing Specification}
\label{subsec:depmodel}

To support the library-aware alias analysis, we concentrate on the API aliasing specification inference
and follow an important form of aliasing specifications formulated in the prior study~\cite{EberhardtSRV19},
which is defined as follows.

\begin{definition}(API Aliasing Specification)
\label{def:spec}
An API aliasing specification is a tuple $(m_1, m_2, P, t)$, where $m_1$ and $m_2$ are two APIs, $P$ := $\{$$(i_1^{(1)}, i_1^{(2)}), \cdots , (i_j^{(1)}, i_j^{(2)})$$\}$ is a set of non-negative integer pairs, and $t$ is an non-negative integer.
It indicates that the return value of $m_2$ can be aliased with the $t$-th parameter of $m_1$ if 
\begin{itemize}[leftmargin=*]
\item $m_1$ is called before $m_2$ upon the same object
\item The $i_k^{(1)}$ and $i_k^{(2)}$-th parameters of $m_1$ and $m_2$ are aliased accordingly.
\end{itemize}
Here, $0 \leq i_k^{(1)} \leq n_1$, $0 \leq i_k^{(2)} \leq n_2$, and $0 \leq k \leq j$.
$n_1$ and $n_2$ are the parameter numbers of $m_1$ and $m_2$, respectively. 
Without ambiguity, we call $m_1$ and $m_2$ a store-load API pair.
\end{definition}

\vspace{-1mm}
Definition~\ref{def:spec} shows that the APIs $m_1$ and $m_2$ conduct the store and load operations upon the memory, respectively.
Unlike simple load and store operations of pointers,
storing and loading the values in memory may depend on the values of other parameters, \revision{which are induced by the set $P$,
determining the memory location where the values are stored and loaded, respectively.}
Essentially, the set $P$ indicates the precondition of the aliasing relation between the return value of $m_2$ and the $t$-th parameter of $m_1$.
If  $P$ is empty, the parameters of $m_1$ and $m_2$ are not necessarily aliased to enforce the aliasing relation between the return value of $m_2$ and the $t$-th parameter of $m_1$.

\begin{example}
\label{ex:2}
In Figure~\ref{fig:javadoc}(a), we have two API aliasing specifications
$(m_1, m_4, \{(0, 0)\}, 1)$ and $(m_2, m_3, \emptyset, 0)$.
Specifically, the API aliasing specification $(m_1, m_4, \{(0, 0)\}, 1)$ indicates that the return value of \textsf{Intent.getStringArrayListExtra} and the second parameter of \textsf{Intent.putStringArrayListExtra} are aliased when they are invoked upon the same object and their first parameters are aliased.
\end{example}

The API aliasing specification in Definition~\ref{def:spec} is more general than the one targeted by \textsc{USpec}~\cite{EberhardtSRV19}.
Specifically, \textsc{USpec} only infers that calling $m_2$ may return a value aliased with the $t$-th parameter of a preceding call of $m_1$ on the same object \emph{if all other parameters are aliased}.
However, there exist many store-load API pairs in which not all the other parameters are aliased.
For instance, the API \textsf{createBitmap} of \textsf{android.graphics.Bitmap} sets the values of \textsf{DisplayMetrics}, \textsf{Config}, width, and height simultaneously, while the \deletion{method}\revision{API} \textsf{getConfig} only fetches the value of \textsf{Config}.
Our formulation in Definition~\ref{def:spec} is expressive enough to depict such a store-load API pair.

\subsection{Problem Statement} 
\label{subsec:ps}

We aim to address the API aliasing specification inference problem from another perspective.
As demonstrated in~\cref{subsec:syntax},
the library documentation provides various forms of semantic information about the library APIs.
Hence, we can hopefully derive the API aliasing specifications from documentation without conducting deep semantic analysis upon the source code or program runtime information.

\revision{The API aliasing specification for a given store-load API pair may not be unique.
In Example~\ref{ex:2}, for instance, $(m_1, m_4, \emptyset, 1)$ is also a valid specification,
while it does not pose any restrictions upon the parameters of the two APIs as the pre-condition.
In our work, we want to ensure that the inferred specifications 
exhibit as strong pre-conditions as possible,
which implies the maximal size of the set $P$.}
Finally, we state the problem of the API aliasing specification inference as follows.

\mybox{
Given a documentation model $\mathbf{L} = (\mathbf{H}, \mathbf{T}, \mathbf{N}, \mathbf{D})$, infer a set of API aliasing specifications $S_{\textsf{AS}}$ such that $|P|$ is maximized for each $(m_1, m_2, P, t) \in S_{\textsf{AS}}$.
}

\emph{\textbf{Technical Challenges.}}
Although library documentation offers semantic information,
solving the above problem is quite challenging.
First, the naming information and semantic descriptions can be ambiguous.
Without an effective interpretation, we cannot understand how the APIs operate on memory or identify aliasing relations between parameters and return values.
Second, there are often many available APIs offered by a single class and even its superclasses.
It is non-trivial to obtain high efficiency in front of a large number of available APIs for each class.

\smallskip
\emph{\textbf{Roadmap.}}
In this work, we propose an inference algorithm \ToolName\ to address the two technical challenges.
Specifically, we introduce the documentation model abstraction 
to formulate semantic information,
which enables us to reduce the original problem to an optimization problem (\cref{sec:abstraction}).
Furthermore, we propose the neurosymbolic optimization to efficiently solve the instantiated optimization problem (\cref{sec:solving}).
\revision{We present the details of our implementation~(\cref{sec:impl}) and demonstrate the evaluation quantifying the effectiveness and efficiency of \ToolName~(\cref{sec:eval}).}\deletion{Our implementation and evaluation demonstrate the effectiveness and efficiency of our approach.}

\section{Documentation Model Abstraction}
\label{sec:abstraction}
This section presents the abstraction of the documentation model.
\revision{Specifically, we propose the concept of the API value graph~(\cref{subsec:graph}) and
introduce two label abstractions over the graph~(\cref{subsec:abstraction}),}
which enables us to reduce the API aliasing specification problem to an optimization problem (\cref{subsec:reduction}).

\subsection{API Value Graph}

As shown in \cref{subsec:syntax}, the formal semantic information, namely class hierarchy and the type signatures, 
reveals potential aliasing relations between API parameters and return values,
while the informal semantic information,
e.g., naming information and semantic descriptions,
shows how parameters and return values are utilized.
To depict aliasing relations that can be introduced by API invocations,
we propose a graph representation, namely \emph{API value graph}, as follows.

\label{subsec:graph}
\begin{definition}[API Value Graph]
\label{def:AVG}
Given a documentation model $\mathbf{L} = (\mathbf{H}, \mathbf{T}, \mathbf{N}, \mathbf{D})$,
its API value graph is the labeled graph 
$G:=(V, E, \ell_n, \ell_d)$, where
\begin{itemize}[leftmargin=*]
    \item The node set $V$ contains API parameters and return values, which are referred to as \emph{API values}.
    $(c, m, i) \in V$ if and only if $(c, m, i) \in \textsf{dom}(\mathbf{N})$ or there is $c' \in \mathbf{H}(c)$ such that $(c', m, i) \in \textsf{dom}(\mathbf{N})$.
    \item The edge set $E \subseteq V \times V$ indicates \revision{possible} aliasing relations between API values.
    Specifically, $(v_1, \ v_2) \in E$ if and only if 
    $\mathbf{T}(v_1) = \mathbf{T}(v_2)$, $\mathbf{T}(v_1) \in \mathbf{H}(\mathbf{T}(v_2))$, or $\mathbf{T}(v_2) \in \mathbf{H}(\mathbf{T}(v_1))$.
    \item The name label $\ell_n$ is a function that maps an API value to its name, i.e., $\ell_n(v) = \mathbf{N}(v)$.
    \item The description label $\ell_d$ is a function that maps an API value to the semantic description of the API, i.e., $\ell_d(v) = \mathbf{D}(c, m)$, where $v=(c, m, i)$.
\end{itemize}
\end{definition}

The API value graph regards API values, 
namely API parameters and return values,
as first-class citizens,
and depicts their high-level semantics with labels.
Intuitively, an edge from $(c, m_1, i_1)$ to $(c, m_2, i_2)$ indicates the fact that the two values may be aliased when $m_2$ is invoked after $m_1$ upon the same object.
Meanwhile, the two labels attach the informal semantic information to API values, showing their usage intention.
From a high-level perspective, the API value graph over-approximates aliasing relations according to class hierarchy relations and type signatures, and still preserves informal semantic information as labels to support further specification inference.

\begin{figure*}[t]
\centering
\includegraphics[width=\linewidth]{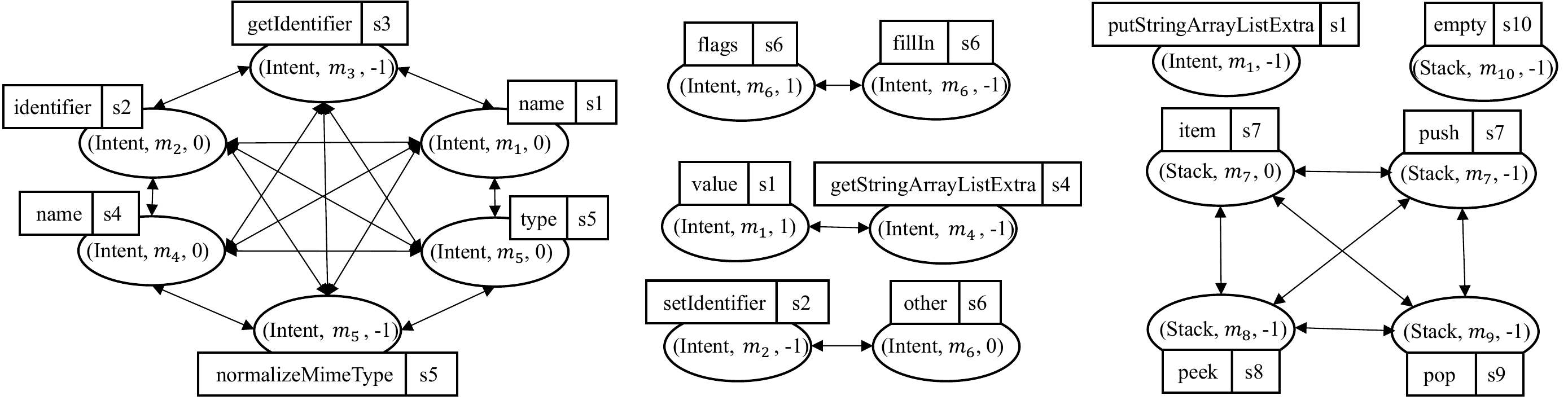}
\vspace{-5mm}
\caption{The API value graph of the documentation model induced by the documentation in Figure~\ref{fig:javadoc}}
\vspace{-2mm}
\label{fig:graph}
\end{figure*}

\begin{example}
Figure~\ref{fig:graph} shows the API value graph for the documentation model induced by the classes in Figure~\ref{fig:javadoc},
where the name labels and description labels are shown in the left and right boxes, respectively.
$s_i$ indicates the semantic description of $m_i$ in Figure~\ref{fig:javadoc}.
Specifically, the edge from $(\textsf{Intent}, m_2, 0)$ to $(\textsf{Intent}, m_5, 0)$ indicates that the first parameters of \textsf{Intent.setIdentifier} and \textsf{Intent.normalizeMimeType} may be aliased when the two APIs are invoked successively.
\end{example}

\subsection{Label Abstraction}
\label{subsec:abstraction}
\revision{Although the edges of the API value graph over-approximate aliasing relations over API values, not all the aliasing relations can hold when using APIs.}
\deletion{The edges of the API value graph approximate aliasing relations over API values based on their types.
However, not all the aliasing relations can hold when using APIs.}
In Figure~\ref{fig:javadoc}, for example, the return value of \textsf{getIdentifier} and the first parameter of \textsf{normalizeMimeType} are unlikely to be aliased
as the named entities in their names are different,
revealing the different usage intention of the two API values.
To formulate this key idea,
we first introduce the concept of the \emph{semantic unit abstraction} as follows.
\deletion{which shows the high-level semantics of API values.}

\begin{definition}(Semantic Unit Abstraction)
\label{def:sua}
\revision{A semantic unit abstraction $\alpha_{\tau}$ is a function mapping a string $s$
to a set of named entities contained in $s$.
We call the elements in $\alpha_{\tau}(s)$ as \emph{semantic units}.}
\deletion{A semantic unit abstraction $\alpha_{\tau}$ maps a string $s$
to a set of nouns, i.e.,
$\alpha_{\tau}(s) = \{w_i  \ | \ \tau(w_i) = \textsf{NOUN}$, $\ s = w_1 \odot w_2 \odot \cdots \odot w_n \}$
Here, $\odot$ is the concatenation operation, and $s$ is the concatenation of $w_i$.
$\tau$ is a tagging function that maps a word to a grammatical tag.
Particularly, we call the named entities in $\alpha_{\tau}(s)$ as the \emph{semantic units} of $s$.}
\end{definition}

\begin{example}
\label{ex:sua}
The named entities in the API name of \textsf{getStringArrayListExtra} include \textsf{string}, \textsf{array}, \textsf{list}, and \textsf{extra}.
Hence, we have
$\alpha_{\tau}(\textsf{getStringArrayListExtra}) = \{
\textsf{string}, \ \textsf{array}, \ \textsf{list}, \ \textsf{extra}\}$.
\end{example}

Essentially, the semantic unit abstraction extracts the named entities from the names as semantic units, which shows the high-level semantics of API values, enabling us to refine aliasing relations according to the following two intuitions:
(1) If two API values $v_1$ and $v_2$ have names with the same semantic units, we can obtain the confidence that they are very likely to indicate the same object in the memory; (2) If the name of an API value does not have any semantic units, we can conservatively regard that it can be aliased with any other API values with consistent types. Hence, we formally define the \emph{semantic unit consistency} to formulate the two intuitions.

\begin{definition}(Semantic Unit Consistency)
\label{def:suc}
Given a semantic unit abstraction $\alpha_{\tau}$ upon an API value graph $G=(V, E, \ell_n, \ell_d)$, two nodes $v_1$ and $v_2$ are semantic-unit consistent if and only if
\deletion{denoted by $(v_1, v_2) \in C_{\tau} \subseteq E$, 
(1) $\alpha_{\tau}(\mathbf{N}(v_1)) = \alpha_{\tau}(\mathbf{N}(v_2))$,
or (2) $\alpha_{\tau}(\mathbf{N}(v_1)) = \emptyset \lor \alpha_{\tau}(\mathbf{N}(v_2)) = \emptyset$. }
\revision{(1) $\alpha_{\tau}(\ell_n(v_1)) = \alpha_{\tau}(\ell_n(v_2))$,
or (2) $\alpha_{\tau}(\ell_n(v_1)) = \emptyset \lor \alpha_{\tau}(\ell_n(v_2)) = \emptyset$. 
}
\end{definition}

\begin{example}
\label{ex:sua}
Consider the API value graph in Figure~\ref{fig:graph}.
We have $\alpha_{\tau}(\textsf{getIdentifier})  = \{ \textsf{identifier} \} $, so the return value of the API \textsf{getIdentifier} and the first parameter of \textsf{setIdentifier} are semantic-unit consistent for the class \textsf{Intent}.
Also, we have $\alpha_{\tau}(\textsf{item}) = \{ \textsf{item} \} $ and $\alpha_{\tau}(\textsf{peek}) = \emptyset $,
so the return value of \textsf{peek} and the first parameter of \textsf{push} are semantic-unit consistent for the class \textsf{Stack}.
\end{example}

\textcolor{black}{Finally, we observe that semantic descriptions characterize how API values interact with memory. By leveraging these informal descriptions, we can systematically identify the underlying memory operations required to infer precise data-flow and aliasing specifications. To formalize this mapping, we define the concept of  \textbf{memory operation abstraction} as follows:}

\begin{definition}(Memory Operation Abstraction)
\label{def:moa}
A memory operation abstraction $\alpha_{o}$ maps a semantic description $s$ to $\alpha_{o}(s) \subseteq M$, where
$M = \{\textsf{I}, \textsf{D}, \textsf{R}, \textsf{W}\}$.
The elements in $M$ indicate the insertion (I), deletion (D), read (R), and write (W) operations on the memory. 
\end{definition}

\revision{
Notably, we classify common memory operations into four categories due to two major reasons.
First, the write operation contains several sub-kinds, such as deletion and insertion. If we only categorize memory operations into read and write, we can not distinguish the APIs conducting the deletion and insertion, such as \textsf{pop} and \textsf{add} for \textsf{java.util.Stack},
which may yield wrong API aliasing specifications.
For example, the APIs \textsf{pop} and \textsf{peek} of \textsf{java.util.Stack} would be wrongly identified to form a store-load pair and thus induce an incorrect API aliasing specification.
Second, objects can be organized in various structural manners. When adding an object to a container-typed field, such as \textsf{java.util.Stack} and \textsf{java.util.HashMap}, the operation is an insertion. When storing an object in a non-container-typed field, the API writes a specific value to the field. 
The above operations are often described differently in natural language,
so we formulate the memory operation abstraction in a fine-grained manner.}

\begin{example}
\label{ex:moa}
According to Figure~\ref{fig:javadoc},
we have $\alpha_{o}(s_1) =  \{\textsf{I}, \ \textsf{W} \}$, $\alpha_{o}(s_2) = \{ \textsf{W} \}$ and
$\alpha_{o}(s_3) = \alpha_{o}(s_4) = \{ \textsf{R} \}$ for \textsf{Intent}.
For \textsf{Stack},
we have $\alpha_{o}(s_7) = \{ \textsf{I}, \textsf{W} \}$, $\alpha_{o}(s_8) = \{\textsf{R}\}$, 
and $\alpha_{o}(s_9) = \{ \textsf{R}, \textsf{D}, \textsf{W} \}$.
\end{example}

To sum up, the above two label abstractions interpret the informal semantic descriptions
with the sets of semantic units and memory operations, based on which we can refine potential aliasing relations indicated by the edges of the API value graph and identify store-load API pairs.
In \cref{subsec:instantiation}, we will demonstrate how to instantiate the two abstractions
to support the specification inference.

\subsection{Problem Reduction}
\label{subsec:reduction}

Based on the two label abstractions,
we can interpret the high-level semantics of API values and the memory operations conducted by the APIs.
According to our problem statement in~\cref{subsec:ps},
we need to identify the store-load API pairs
\revision{and find as many aliased parameters as possible, which determine a strong precondition of the aliasing relation between loaded and stored values.}
\deletion{Besides, we need to infer as many aliased parameters in each API pair as possible so that the inferred specification can pose a strong precondition over the API parameters, which finally induces a precise abstraction of the API semantics.}
Hence, we reduce the specification inference to an optimization problem over the API value graph as follows.

\begin{definition}(Optimization Problem)
\label{def:opt}
Given a semantic unit abstraction $\alpha_{\tau}$ and a memory operation abstraction $\alpha_{o}$ upon an API value graph $G = (V, E, \ell_n, \ell_d)$,
find an edge set $E^{*} \subseteq E$ with a maximal size $|E^{*}|$ satisfying the following constraints: 

\begin{itemize}[leftmargin=*]
    \item (Degree constraint) \revision{For each $v \in V$, the in-degree and out-degree of $v$ are not greater than 1. }
    \deletion{For any $v=(c, m, i) \in V$ and $m'$, the following two conditions are satisfied}
    \item (Validity constraint) If $(v_1, v_2) \in E^{*}$, where $v_1$ and $v_2$ \revision{indicate parameters}, there exist $u_1, \ u_2 \in V$ such that $(u_1, u_2) \in E^{*}$.
    where $u_1$ and $u_2$ indicate a parameter and a return value, respectively.
    
    \item (Semantic unit constraint) For any $(v_1, v_2) \in E^{*}$, where $v_1 = (c, m_1, i_1)$ and $v_2 = (c, m_2, i_2)$, the semantic unit abstraction of the names of $v_1$ and $v_2$ should satisfy
    \begin{itemize}
        \item \revision{(S1)} If $i_2 \neq -1$, $v_1$ and $v_2$ are semantic-unit consistent.\deletion{, i.e., $(v_1, v_2) \in C_{\tau}$}
        \item \revision{(S2)} If $i_2 = -1$, $v_1$ or $v_1'$ is semantic-unit consistent with $v_2$, where $v_1' = (c, m_1, -1)$.\deletion{, i.e., $(v_1, v_2) \in C_{\tau}$ or $(v_1', v_2) \in C_{\tau}$.}
    \end{itemize}

    \item (Memory operation constraint) For any $(v_1, v_2) \in E^{*}$, the following two conditions are satisfied:
    \begin{itemize}
        \item \revision{(M1)} $v_1$ satisfies $ \textsf{I} \in \alpha_{o}(\ell_d(v_1)) \lor (\textsf{W} \in \alpha_{o}(\ell_d(v_1))\land \textsf{D} \notin  \alpha_{o}(\ell_d(v_1))$
        \item \revision{(M2)} $v_2$ satisfies that $\textsf{R} \in \alpha_{o}(\ell_d(v_2))$
    \end{itemize}    
\end{itemize}
\end{definition}

Definition~\ref{def:opt} aims to maximize $| E^{*} |$ to discover all the aliased parameters of each store-load API pair\revision{, which corresponds to maximizing $|P|$ in original problem statement in~\cref{subsec:ps}.}
The four kinds of constraints are imposed upon the selected edges. Specifically, the degree and validity constraints ensure that the edges induce the API aliasing specification defined in Definition~\ref{def:spec}.
Besides, the parameters of the APIs $m_1$ and $m_2$ should be semantic-unit consistent if they are connected by a selected edge \revision{(S1)}.
If a selected edge connects the parameter of $m_1$ and the return value of $m_2$,
then the parameter of $m_1$ should be semantic-unit consistent with the return value of $m_2$ \revision{(S2)}.
Lastly, the memory operation constraint ensures that the APIs $m_1$  and $m_2$ should form a store-load API pair \revision{(M1 and M2)}. 
Finally, we can obtain the specifications based on the optimal solution as follows.

\mybox{
	Given the optimal solution $E^{*}$ of the optimization problem defined in Definition~\ref{def:opt},
	we can obtain the API aliasing specification $(m_1, m_2, P, t) \in S_{\textsf{AS}}$, where
	\begin{itemize}[leftmargin=*]
		\item $P = \{ (i_1, i_2) \ | \ ((c, m_1, i_1), \ (c, m_2, i_2)) \in E^{*}, i_2 \neq -1  \}$
		\item $t$ satisfies $ ((c, m_1, t), (c, m_2, -1)) \in E^{*} $
	\end{itemize}
}

\begin{figure}[t]
  \centering
\includegraphics[width=0.8\linewidth]{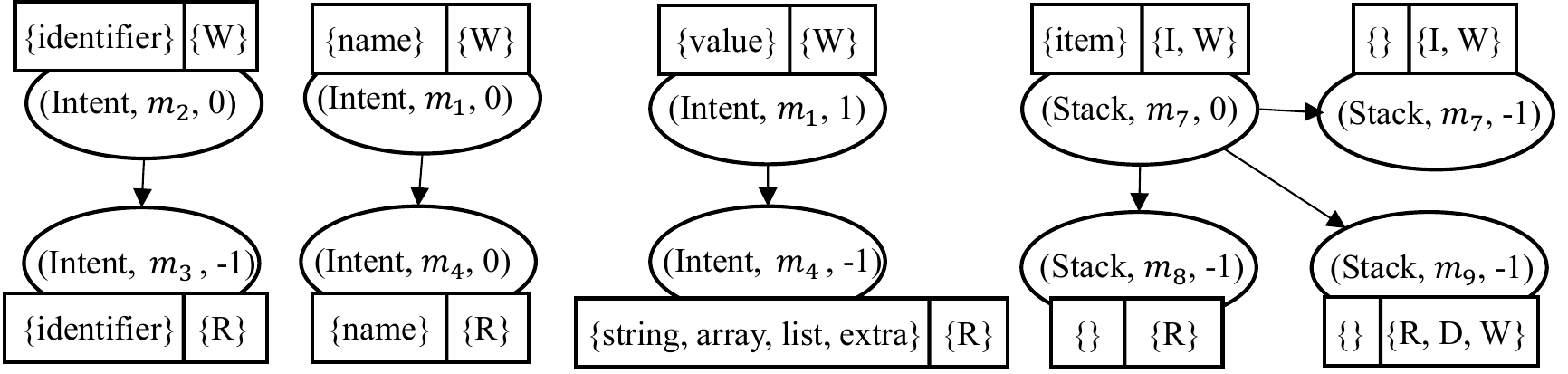}
  \caption{An optimal solution to the problem instance over the API value graph shown in Figure~\ref{fig:graph}}
  \label{fig:pairing}
\vspace{-6mm}
\end{figure}

\begin{example}
Figure~\ref{fig:pairing} shows the optimal solution to the optimization problem over the API value graph in Figure~\ref{fig:graph}, where the sets shown in the two boxes demonstrate the extracted semantic units and the identified memory operations under the label abstractions in Examples~\ref{ex:sua} and~\ref{ex:moa}.
We discover six possible aliasing relations.
Notably, although the semantic units of $(\textsf{Intent}, m_4, -1)$ are different from $(\textsf{Intent}, m_1, 1)$, they are exactly the same as the ones of $(\textsf{Intent}, m_1, -1)$, indicating that the second parameter of $m_1$ can have the same semantics as the return value of $m_4$.
The optimal solution finally induces the API aliasing specifications in Example~\ref{ex:2}.
\end{example}

By reducing the original problem to the optimization problem in Definition~\ref{def:opt},
we only need to tackle two sub-problems for the specification inference.
First, we have to instantiate two label abstractions to precisely interpret the semantic meanings of names and the kind of memory operations.
Second, we need to design an efficient optimization algorithm to solve the optimization problem.\deletion{ and further convert the optimal solution to the API aliasing specifications.}
In \cref{sec:solving}, we will provide the technical details of addressing the two sub-problems.
\section{Specification Inference via Neurosymbolic Optimization}
\label{sec:solving}

This section presents the technical details of our algorithm \ToolName.
Specifically, we demonstrate the overall algorithm in \cref{subsec:overallalg} and detail the label abstraction instantiation in \cref{subsec:instantiation}.
Besides, we present the neurosymbolic optimization in \cref{subsec:neuroopt} to instantiate and solve the optimization problem given in Definition~\ref{def:opt}.
Lastly, we summarize our approach and highlight its advantages in \cref{subsec:summary}.

\subsection{Overall Algorithm}
\label{subsec:overallalg}

\begin{wrapfigure}[11]{t}{0.5\textwidth}
	\centering
	\vspace{-5mm}
	\scalebox{0.92}{
		\begin{minipage}{0.52\textwidth}
			\centering
			\begin{algorithm}[H]
				\SetNoFillComment
				\caption{Inference Algorithm}
				\label{alg:overallalg}
				\KwIn{$\mathbf{L}$: Documentation model;}
				\KwOut{$S_{\textsf{AS}}$: API aliasing specifications;}
				\SetKwFunction{constructAPIValueGraph}{constructAVG}
				\SetKwFunction{instantiateSemanticUnitAbstraction}{getSemanticUnitAbs}
				\SetKwFunction{instantiateMemoryOperationAbstraction}{getMemoryOperationAbs}
				\SetKwFunction{neuroSymOpt}{neuroSymOpt}
				\SetKwFunction{convert}{convert}
				$G \leftarrow \constructAPIValueGraph(\mathbf{L})$\;
				$\alpha_{\tau} \leftarrow \instantiateSemanticUnitAbstraction()$\;
				$\alpha_{o} \leftarrow \instantiateMemoryOperationAbstraction()$\;
				$\mathcal{P} \leftarrow (\mathbf{L}, G, \alpha_{\tau}, \alpha_{o})$\;
				$E^{*} \leftarrow \neuroSymOpt(\mathcal{P})$\;
				$S_{\textsf{AS}} \leftarrow \convert(E^{*})$\;
				\Return{$S_{\textsf{AS}}$}\;
			\end{algorithm}
		\end{minipage}
	}
\end{wrapfigure}

As demonstrated in \cref{subsec:reduction}, we can reduce the API aliasing specification inference problem to an instance of the optimization problem given in Definition~\ref{def:opt}.
Technically, we propose and formulate our specification algorithm in Algorithm~\ref{alg:overallalg},
which takes as input a documentation model $\mathbf{L}$ and 
generates a set of API aliasing specifications $S_{\textsf{AS}}$ as output.
First, we derive the API value graph $G$ from the documentation model $\mathbf{L}$ based on Definition~\ref{def:AVG} (Line~1).
Second, we instantiate two label abstractions, i.e., $\alpha_{\tau}$ and $\alpha_{o}$, and further construct an instance of the optimization problem $\mathcal{P}$ defined in Definition~\ref{def:opt} (Lines 2--3).
Third, we propose the neurosymbolic optimization to solve the instance of the optimization problem $\mathcal{P}$ (Lines 4--5),
and finally convert the optimal solution $E^{*}$ to a set of API aliasing specifications $S_{\textsf{AS}}$ (Line 6).
Particularly, Definition~\ref{def:AVG} has demonstrated how to construct the API value graph,
and converting the optimal solution to the specification is also explicitly formulated at the end of \cref{subsec:reduction}.
In the rest of this section, we will provide more details on the label abstraction instantiation (\cref{subsec:instantiation}) and the neurosymbolic optimization algorithm (\cref{subsec:neuroopt}),
which finalize the functions \textsf{getSemanticUnitAbs}, \textsf{getMemoryOperationAbs}, and \textsf{neuroSymOpt} in Algorithm~\ref{alg:overallalg}, respectively.

\subsection{Label Abstraction Instantiation}
\label{subsec:instantiation}
According to Definitions~\ref{def:sua} and~\ref{def:moa}, the semantic unit abstraction requires attaching the grammatical tags,
while the memory operation abstraction demands identifying how an API manipulates memory.
In what follows, we will detail how to instantiate them with two different NLP models, respectively.


\subsubsection{Instantiating Semantic Unit Abstraction.}
According to common programming practices, the developers of libraries tend to follow typical naming conventions~\cite{ButlerWY15}, such as camel case, pascal case, and snake case.
For example, \textsf{userAccount} is a parameter name using camel case, and \textsf{get\_account\_balance} is an API name using snake case.
Notably, the sub-words are often separated with an underscore or begin with an uppercase letter.
Hence, we can easily decompose each name $s$ into the concatenation of several sub-words and further determine the tag of each sub-word.

However, the names of APIs or their parameters can hardly be valid phrases or sentences.
Simply applying the part-of-speech (POS) tagging would tag almost all the words as nouns.
\revision{Also, the POS tagging targets tagging sentences, while the names of parameters and APIs are only the concatenation of words in phrases.}
To obtain more precise tagging results, we leverage an existing probability model trained in Brown Corpus~\cite{francis1967computational}, which can return all the possible grammatical tags of each word along with the occurrences.
\revision{This enables us to determine whether a word is more likely to be a noun according to the existing probability model,
which does not depend on the usage context of the word.}
Formally, we instantiate the semantic unit abstraction as follows.

\begin{definition}(Instantiation of Semantic Unit Abstraction)
Assume that $g_{\tau}$ maps a word $w$ to a set of tag-occurrence pairs $\{ (\tau_j, k_j) \}$.
\revision{Given a sub-word $w$ in a parameter/API name $s$, $w \in \alpha_{\tau}(s)$ if and only if $(\textsf{NOUN}, k^{*}) \in g_{\tau}(w)$
and $k^{*}$ is the largest occurrence in $g_{\tau}(w)$.}
\deletion{Then, the tagging function $\tau$ is defined as
$\tau(w) = \mathop{\arg\max}_{\tau_j} \textsf{o}(\tau_j)$,
where $(\tau_j, \textsf{o}(\tau_j)) \in g_{\tau}(w)$,
which further instantiates $\alpha_{\tau}$.}
\end{definition}

\begin{example}
Consider the API \textsf{setIdentifier} in Figure~\ref{fig:javadoc}.
After splitting the API name into two sub-words, namely ``set\text{''} and ``identifier\text{''},
we discover that ``set\text{''} is more likely to be a verb than a noun,
while ``identifier\text{''} is very likely to be a noun.
Hence, our instantiated semantic unit abstraction $\alpha_{\tau}$ maps \textsf{setIdentifier} to $\{ \textsf{identifier} \}$,
identifying \textsf{identifier} as the semantic unit of the API.
\end{example}

\subsubsection{Instantiating Memory Operation Abstraction.}
\label{subsubsec:moa_instantation}

NLP models are particularly effective at distilling program semantics from natural language descriptors, enabling the autonomous inference of API specifications - such as taint specifications \cite{fluffy,docflow,binarytaint10.1145/3711816} and alias specifications \cite{DAInfer} - from unstructured documentation.
Human-written API descriptions serve as high-level functional specifications, providing valuable semantic information regarding a method’s behavioral intent.

To instantiate an effective memory abstraction, we leverage two common programming practices: (1) developers typically summarize API functionality using full sentences or verb-object phrases as semantic descriptions, and (2) the verbs within these descriptions intuitively depict the underlying memory operations performed by the API.
However, extracting these operations is hindered by the vocabulary mismatch problem; for example, the verbs ``\textsf{put}'', ``\textsf{insert}'', and ``\textsf{push}'' may all denote a single memory insertion primitive. 
The diverse choices of the verbs describing a specific memory operation would make the inference suffer low recall if we just adopted a grep-like approach based on string matching.

\begin{figure*}[t]
	\centering
	\includegraphics[width=\linewidth]{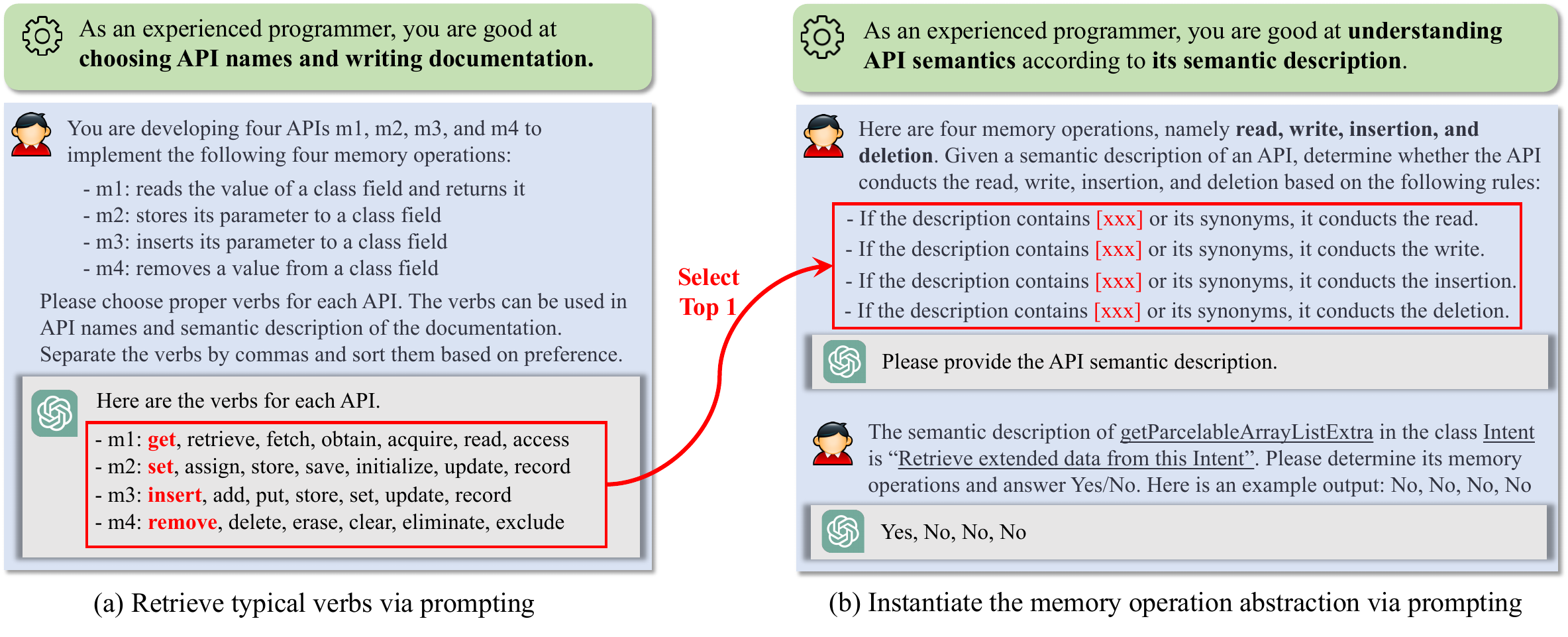}
	\vspace{-7mm}
	\caption{\revision{The prompt templates of two-staged prompting}\deletion{Instantiate the memory operation abstraction via two-staged prompting}}
		\vspace{-5mm}
	\label{fig:prompt}
	\vspace{4mm}
\end{figure*}

Initially, we are inspired by recent progress in the NLP community. Hence, we realize that the latest advances in the LLMs may provide new opportunities for resolving this issue~\cite{chatgpt, openai2023gpt4, DBLP:conf/nips/BrownMRSKDNSSAA20}.
Specifically, the LLMs have excellent abilities in text understanding, especially under the guidance of a few-shot examples or descriptions of rules. 
We designed a two-staged prompting solution that infers the memory operation kind, considering the best practices followed by developers in selecting the name for an API method, given the API description.
While using LLMs to reason over descriptions helps, they are prone to hallucinations, occasionally misclassifying intent or inventing non-existent operations in linguistic structures. For instance, the description ``\textbf{Removes the object at the top of this stack and returns that object as
the value of this function}'' for API method ``\textbf{pop()}'' consists of two simple sentences connected with the connecting word ``\textbf{and}''. The verb ``\textbf{remove}'' implies a modification on the memory while the verb ``\textbf{return}'' shows a memory read operation afterwards. Although the prompt shown in Fig. \ref{fig:prompt} requests LLM to choose all the relevant semantic descriptions, it is possible that it is biased by one part of the complex sentence, not producing the desired output.

\textcolor{black}{In this extended version, we propose a more robust solution using embedding models. 
By mapping API descriptions and formal verb-phrases describing memory primitives into a shared high-dimensional vector space, we can perform semantic similarity comparison. 
This deterministic approach resolves vocabulary variance while providing a grounded safeguard against the hallucinations inherent in purely generative models.}

\textcolor{black}{\textbf{Memory Abstraction with LLMs: }}
\textcolor{black}{To} instantiate the memory operation abstraction, \textcolor{black}{we propose} two-stage prompting,
of which the prompt template is shown in Figure~\ref{fig:prompt}.
\begin{itemize}[leftmargin=*]
\item First, we design the prompt in Figure~\ref{fig:prompt} (a) to retrieve the verbs describing each memory operation and enforce the LLMs sort them based on the preference.
\revision{Although the verb lists may overlap,
the top-1 verbs are representative enough to distinguish different memory operations.}
\item Second, we select the top-1 verbs recommended in the first stage and then construct the prompt describing the rules for the memory operation abstraction, which is shown in Figure~\ref{fig:prompt}(b).
Finally, we obtain an LLM response containing four ``Yes\text{''}/``No\text{''} separated by commas.
\end{itemize}

\revision{
It is worth noting that we identify memory operation kinds via a two-stage prompting instead of a one-stage prompting. If we manually specify the typical verbs describing memory operations, the second prompt may rely on our manual setting, which demands expert knowledge. If we do not offer typical words as hints, the result is not as interpretable as the current one. Our design actually utilizes the ability of LLMs to predict method names for coding tasks, self-promoting the memory operation identification with generated typical verbs. Note that the first stage is only conducted once. The typical verbs are shared when analyzing library APIs. Hence, the extra cost introduced by the first stage is negligible.
Based on the above prompting process, we can obtain an instantiation of the memory operation abstraction, which is formally formulated as follows.}

\begin{definition}(Instantiation of Memory Operation Abstraction)
$g_{o}$ is the function induced by the LLM via two-staged prompting in Figure~\ref{fig:prompt}.
Then the memory operation abstraction $\alpha_{o}$ satisfies that $\textsf{op} \in \alpha_{o}(s)$ if and only if the corresponding answer of $\textsf{op}$ in $g_{o}(s)$ is ``Yes\text{''}, where $\textsf{op} \in M$.
\end{definition}

\begin{example}
In Figure~\ref{fig:prompt}(b), the output of the LLM is ``Yes, No, No, No\text{''}, indicating that \textsf{Intent.getStringArrayListExtra} only conducts the memory read. Hence, we have $\alpha_{o}(s_4) = \{ \textsf{R} \}$, where $s_4$ is the semantic description of the API \textsf{Intent.getStringArrayListExtra}.
\revision{
Similarly, for the API \textsf{Intent.normalizeMineType}, the verb ``normalize\text{''} in its semantic description $s_5$ is not the synonym of four typical verbs, so $\alpha_{o}(s_5) = \emptyset$,
indicating that it does not contribute to any load-store match.
}
\end{example}

\textbf{Memory Abstraction with Embedding Models:} By representing both API descriptions and memory operation descriptions as vectors, we can quantify their semantic similarity and identify the most likely memory behaviors (e.g., read, write, insert, or remove operations) without relying on rigid keyword matching. A semantic description $s$ is an informal description of an API method's functionality through a set of sentences. To reliably infer the memory operation from a semantic description, it is essential to consider the overall semantics of all the sentences. However, interpreting the connection between the sentences when they are not simple is not straightforward. 

\begin{figure}[!ht]
    \centering
    \resizebox{\textwidth}{!}{
    \begin{tabular}{ccc}
      \begin{forest}
        [Simple Sentence (S)
          [Verb Phrase (VP)
             [Verb [Sets]]
			 [Noun Phrase (NP) [an identifier]]
          ]
        ]
\end{forest}
        & 
       \begin{forest}
            [Compound Sentence (S)
              [Independent Clause ($sent_1$)
                [VP [Remove the object ...]]
              ]
              [Coordinator [and]]
              [Independent Clause ($sent_2$)
                [VP [Return the object]]
              ]
            ]
            \end{forest}
        & 
     \begin{forest}
        [Complex Sentence (S)
          [Main Clause
            [VP [Tests]]
          ]
          [Dependent Clause ($sent_{\text{dep}}$)
            [Subordinator [if]]
            [Clause
              [NP [this stack]]
              [VP [is empty]]
            ]
          ]
        ]
        \end{forest}\\
        (a) Simple & (b) Compound & (c) Complex
    \end{tabular}
    }
    \caption{Visual representation of sentence structures: (a) Simple, (b) Compound, and (c) Complex.}
    \label{fig:sentence_structures}
\end{figure}
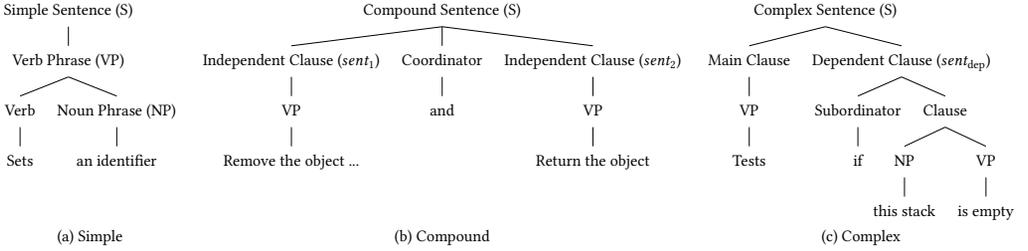

A sentence can have three different structures according to linguistic resources \cite{Backman2004Building} shown in Figure \ref{fig:sentence_structures}. A simple sentence is only an independent clause consisting of a verb phrase and a subject.
A compound sentence consists of two or more independent clauses joined by a coordinating conjunction (FANBOYS: For, And, Nor, But, Or, Yet, So) or a semicolon. A complex sentence includes one independent clause with at least one dependent clause (starts with a subordinating conjunction like when, because, or although). While a simple sentence typically implies a direct functional objective, the latter two structures often provide additional context or conditions regarding the API's behavior.

In a compound sentence, coordinating conjunctions link multiple independent clauses, typically representing a sequence or a set of distinct actions performed by the API. In contrast, in a complex sentence, the independent clause generally denotes the primary action, while the dependent clause specifies the conditions, constraints, or consequences. Consequently, we design a solution that identifies the sentence structure of an API description and employs an induction approach to isolate and infer abstract memory operations by focusing exclusively on the described primary action. This process follows three steps:

\begin{itemize}
    \item \textbf{Step 1: Semantic Structural Decomposition.} We first analyze the semantic description. If it is either a compound or complex sentence, we decompose it into a set of constituent simple sentences \(\{sent_{1},\dots, sent_{n}\}\). 
    \item \textbf{Step 2: Memory Operation Abstraction Inference via Semantic Similarity.} For each extracted sentence $sent_i$, we determine the most likely abstract memory operation $op^{*}_i \in M$ by computing the semantic similarity between its vector representation and the embeddings of our predefined memory operation descriptions. We then select the operation that yields the highest similarity score.
    \item \textbf{Step 3: Memory Operation Abstraction Aggregation.} Finally, we aggregate the inferred operations according to the original sentence structure to finalize the set of abstract memory operations $OP^*$ that characterize the data-flow behavior of the API method.
\end{itemize}

\begin{definition}
    \textcolor{black}{Sentence Structure: For a given semantic description \(s\), we define a typing function called as \(\mathcal{T}(s)\in \{\text{Simple,\ Compound,\ Complex}\}\) that maps the description to its primary linguistic structure based on the connectivity of its constituent clauses.} 
\end{definition}

\begin{definition}
    \textcolor{black}{Semantic Structural Decomposition: A semantic description \(s\) is formally represented as the union of \(n\) simple sentences \(\{sent_{1},\dots ,sent_{n}\}\), where \(n>0\). Each \(sent_i\) represents a fundamental unit of the API method's functionality. }
\end{definition}

\begin{example}
The API documentation for the method \textsf{Stack.pop()} describes it as: ``\textsf{Removes the object at the top of this stack and returns that object as the value of this function}.'' This description uses a compound sentence structure joined by the conjunction ``and.'' We decompose this into two simple sentences:
\begin{itemize}
    \item $sent_1$: ``\textsf{Removes the object at the top of this stack.}''
    \item $sent_2$: ``\textsf{Returns that object as the value of this function.}''
\end{itemize}
\end{example}

\begin{definition} \textcolor{black}{Abstract Memory Operation Descriptions: The abstraction \(\beta \) maps each memory operation \(op \in M=\{R,W,I,D\}\) to a natural language semantic descriptor \(d\in Desc\). The mapping is defined as:}
\begin{itemize}
    \item  \textbf{$\beta(R)$}: Gets value of something.
    \item  \textbf{$\beta(W)$}: Sets value of something.
    \item  \textbf{$\beta(I)$}: Inserts something into a collection.
    \item  \textbf{$\beta(D)$}: Removes something from a collection.
\end{itemize}
\end{definition}

\begin{definition}
Sentence Similarity Function: The similarity function \(f:Sent\times Desc\rightarrow [0,1]\) evaluates the semantic alignment between the vector representation of a sentence $sent_i$ and the vector representation of an operation description $\beta(op)$: \[ score = f(sent_i, \beta(op)),\  op \in M \]
\end{definition}
For each simple sentence, the underlying memory operation is assumed to be encoded within its verb phrases. We map these linguistic units to the set of operations \(M=\{I, D, R, W\}\) by identifying the operation \(op \in M\) whose description \(\beta (op)\) most closely resembles the action of the sentence. Formally, for \(sent_{i}\), we select \(op_{i}^{*}\) so that it maximizes the similarity score \(f\).

\begin{definition}
Sentence-level Memory Operation Abstraction:  The inferred operation $op_i^*$ for $sent_i$ is defined as the operation that maximizes the similarity score:
 \[ op^{*}_{i} = arg\ max_{op \in M}\ f(sent_i, \beta(op))\]
\end{definition}
Once the memory operation abstraction is retrieved for each simple sentence, we attempt to infer the memory operation for the semantic description $s$ provided for an API method. 

\begin{example}
Using a sentence embedding model (e.g., \textsc{SBERT-MPNet}), we calculate the cosine similarity scores between the simple sentence $sent_1$ ``\textsf{Removes the object at the top of this stack}'' and each of the four memory abstraction descriptions as bellow:
\begin{itemize}
    \item $score_r = f(sent_1, \beta(R)) = 0.31$ 
    \item $score_w = f(sent_1, \beta(W)) = 0.24$     
    \item $score_i = f(sent_1, \beta(I)) = 0.29$    
    \item $score_d = f(sent_1, \beta(D)) = 0.59$
\end{itemize}
The optimal operation $op_1^*$ is determined by selecting the abstraction with the highest similarity score:
\[ op_1^* = \text{arg max}_{k \in \{R, W, I, D\}} (score_k) = D \]

\end{example}

\begin{definition}
    \textcolor{black}{Instantiation of Memory Abstraction Operation: Let a semantic description \(s\) be a sentence composed of \(n\) constituent simple sentences \(\{sent_{1},\dots ,sent_{n}\}\). The Induction Function \(\mathcal{G}\) aggregates the individual operation abstractions \(op_{i}^{*}\) into a global operation \(OP^{*}\) based on the sentence structure $\mathcal{T}(s)$ as follows:}
    \[
    OP^* = \mathcal{G}(s, \{op^*_i\}_{i=1}^n) = 
    \begin{cases} 
        \{op^*_1\} & \text{if } \mathcal{T}(s) = \text{Simple} \\
        \bigcup_{i=1}^n \{op^*_i\} & \text{if } \mathcal{T}(s) = \text{Compound} \\
        \{op^*_{\text{independent}}\} & \text{if } \mathcal{T}(s) = \text{Complex}
    \end{cases}
    \]
\end{definition}
Based on the sentence structure of this description, we interpret the memory operation differently with the induction function \(\mathcal{G}\) as stated below:
 \begin{itemize}
    \item Simple Sentence: For \(n=1\), the global operation is directly mapped from the single constituent sentence.
    \item Compound Sentence: The global operation is the union of operations from all clauses, representing a sequence or concurrent set of actions.
    \item Complex Sentence: The global operation is inherited exclusively from the independent clause, which identifies the primary functional intent, while dependent clauses (providing conditions or constraints) are disregarded.
\end{itemize}

By applying \(\mathcal{G}\), we transform unstructured natural language into a structured set of abstract memory operations that can be directly mapped to the nodes of the API value graph.

\revision{Notably, our intuition of the label abstraction upon the API value graph is applicable for general libraries in real-world production. Typically, the developers of libraries are often in well-organized communities and cooperatives, following good naming conventions and using proper verbs in semantic descriptions. 
That is, they are unlikely to use different nouns to indicate the objects with the same usage intention or describe memory operations conducted by the APIs with wrong verbs. Their good development habits permit us to correctly interpret the informal semantic properties of library APIs with the \textcolor{black}{NLP models}, which can yield satisfactory precision and recall in the wild. Our evaluation also demonstrates the effectiveness of the label abstraction upon benchmarks used in existing studies~\cite{Bastani0AL18, EberhardtSRV19, Arzt14FlowDroid}.
Furthermore, such well-structured natural language descriptions, including documentation and comments, have been utilized in various software engineering tasks, such as \textcolor{black}{API name recommendation} \cite{clear,APIGen,methodNameRecommendation1,prema}, API misuse detection~\cite{DBLP:conf/kbse/ZhongZXM09, DBLP:conf/kbse/RenYXXXZS20} and unit test generation~\cite{DBLP:conf/issta/BlasiGKGEPC18}. These approaches, which share similar assumptions about natural language descriptions as ours, have demonstrated their practical impacts in understanding code semantics and benefiting downstream clients. We will provide a detailed discussion of these approaches in~\cref{sec:related}.
}

\subsection{Neurosymbolic Optimization}
\label{subsec:neuroopt}
\revision{As shown in~\cref{subsec:instantiation}, our two label abstractions are achieved with different overheads.
Specifically, the semantic unit abstraction only relies on the tagging model that can be applied efficiently. 
\textcolor{black}{For the memory operation abstraction, we utilize embedding models—a significant optimization over the full LLM inference used in previous iterations—as they offer a better balance between semantic accuracy and computational cost.
}
To achieve high efficiency, we propose a solving technique, named \emph{neurosymbolic optimization}, for the optimization problem defined in Definition~\ref{def:opt}.} 

For each API pair,
we first check the satisfiability of degree constraint $\phi_d$ and the validity constraint $\phi_v$ (Lines 2--5).
If both of them are satisfied, we apply the tagging model to derive the semantic unit constraint $\phi_s$ (Line 6) and examine the satisfiability of the conjunction of the three constraints (Line 7).
If it is satisfiable, \textcolor{black}{we invoke the embedding models} to achieve the memory operation abstraction, and derive the memory operation constraint (Line~9).
Based on OMT solving~\cite{BjornerPF15}, we select the maximal number of edges connecting the API values (Line 10)
and append them to the set $E^{*}$ (Line 11), which is returned as the solution to the optimization problem.

\begin{wrapfigure}[16]{r}{0.49\textwidth}
	\vspace{-5mm}
	\centering
	\scalebox{0.9}{
		\begin{minipage}{0.54\textwidth}
			\centering
			\begin{algorithm}[H]
				\SetNoFillComment
				\caption{Neurosymbolic optimization}
				\label{alg:solving}
				\KwIn{$\mathcal{P}$: An optimization problem;}
				\KwOut{$E^{*}$: The optimal solution;}
				\SetKwFunction{constructAPIValueGraph}{constructAPIValueGraph}
				\SetKwFunction{getSemanticUnitAbstraction}{getSemanticUnitAbstraction}
				\SetKwFunction{getMemoryOperationAbstraction}{getMemoryOperationAbstraction}
				\SetKwFunction{SMTSolve}{SMTSolve}
				\SetKwFunction{OMTSolve}{Solve}
				\SetKwFunction{deriveDegreeConstraints}{deriveDegreeConstraints}
				\SetKwFunction{deriveValidityConstraints}{deriveValidityConstraints}
				\SetKwFunction{deriveSUConstraints}{deriveSUConstraints}
				\SetKwFunction{deriveMOConstraints}{deriveMOConstraints}
				\SetKwFunction{obj}{obj}
				\ForEach{$(c, m_1)$, $(c, m_2)$}{
					$\phi_{d} \leftarrow \deriveDegreeConstraints(\mathcal{P})$\;
					$\phi_{v} \leftarrow \deriveValidityConstraints(\mathcal{P})$\;
					\If{\SMTSolve($\phi_{d} \land \phi_{v}$)=\textsf{UNSAT}}{
						\Continue\;
					}
					$\phi_{s} \leftarrow \deriveSUConstraints(\mathcal{P})$\;
					\If{\SMTSolve($\phi_{d} \land \phi_{v} \land \phi_{s}$)=\textsf{UNSAT}}{
						\Continue\;
					}
					$\phi_{o} \leftarrow \deriveMOConstraints(\mathcal{P})$\;
					$E' \leftarrow \OMTSolve(\obj(\mathcal{P}), \phi_{d} \land \phi_{v} \land \phi_{s} \land \phi_{0})$\;
					$E^{*} \leftarrow E^{*} \cup E'$\;
				}
				\Return{$E^{*}$}\;
			\end{algorithm}
	\end{minipage}	}
    \vspace{-8mm}
\end{wrapfigure}

\revision{
Notably, the degree constraint and validity constraint do not depend on any NLP models and are instantiated \emph{symbolically},
while the semantic unit constraint and memory operation constraint rely on the outputs of the tagging model and \textcolor{black}{ the embedding models}, respectively, being instantiated in a \emph{neural} manner.}
\revision{By decoupling the \emph{symbolic constraints} from \emph{neural ones},
\ToolName\ applies NLP models with a lazy strategy.
Note that in our previous attempt, LLM inference consumed much more time than SMT solving~\cite{DAInfer}.
Our \textcolor{black}{new design with embedding models can significantly reduce the time overhead and the hardware cost.}}

\begin{example}
Consider the APIs of \textsf{Intent} in Figure~\ref{fig:javadoc}(a). When processing the APIs \textsf{Intent.fillIn} and \textsf{Intent.getIdentifier}, the validity constraint is not satisfied as there are no type-consistent parameters or return values. Hence, we do not apply the tagging model or the \textcolor{black}{embedding model}. For the APIs \textsf{Intent.setIdentifier} and \textsf{Intent.normalizeMimeType}, we find that their parameters and return values are not semantic-unit consistent, so we do not invoke the embedding model with their semantic descriptions.
\end{example}

\revision{
	When we designed the label abstraction instantiation,
	we also considered directly prompting LLMs to validate semantic unit consistency.
	However, pairwise examining the names of an API and its parameters introduces a large number of LLM inferences, which increases time and token costs.
	We add more discussions on the possibility of utilizing LLMs to improve \ToolName\ \textcolor{black}{in this task} in~\cref{subsec:discussion}.
}
\vspace{-1mm}
\subsection{Summary}
\label{subsec:summary}

\ToolName\ is the first trial of inferring API specifications from documentation. It demonstrates the promising potential of utilizing new advances in the community of natural language processing, especially \textcolor{black}{embedding models, to solve traditional static analysis problems.}
\revision{Similar to traditional pointer analyses upon source code,
such as Andersen-style pointer analysis~\cite{andersen1994program},
\ToolName\ establishes a constraint system over library documentation to pose restrictions upon pointer facts.
To precisely understand the natural language,
it utilizes the NLP models as documentation interpreters to abstract informal semantic information,
which supports instantiating an optimization problem for the specification inference.
Our insight into utilizing NLP models for documentation interpretation can be generalized in other tasks, such as program synthesis~\cite{10.1145/3452296.3472910} and test case generation~\cite{DBLP:conf/icse/MotwaniB19}.}

\section{Implementation}
\label{sec:impl}
We implement the approach \ToolName\ as a prototype and release the source code online~\cite{DAInferplusRepo}.
Specifically, we implement the documentation parser by using the \emph{BeautifulSoup} Python package.
For each documentation page describing the API semantics,
we can extract four kinds of information,
including class hierarchy relations, API type information, naming information, and API semantic descriptions.
Since library documentation pages almost have a uniform format,
we do not have to make major changes to the implementation of the parser to adapt to libraries.
To instantiate the semantic unit abstraction, we utilize the conditional frequency distributions tool with the Brown Corpus provided by \textsf{the Natural Language Toolkit}~\cite{NLTK}. This allows us to determine whether a word is most likely to be a noun. Finally, we leverage an advanced NLP library called \emph{spaCy} \cite{spacy} to construct the dependency trees, enabling the extraction of clausal structures from simple, compound, and complex sentences.
We utilize the conditional frequency distributions tool with the Brown Corpus provided by \textsf{the Natural Language Toolkit}~\cite{NLTK} to determine whether a word is most likely to be a noun. \textcolor{black}{Finally, we leverage an advanced NLP library called \textsf{spaCy}~\cite{spacy} to construct the dependency trees, enabling the extraction of clausal structures from simple, compound, and complex sentences}.
To instantiate the memory operation abstraction \textcolor{black}{ as in our previous research using LLMs~\cite{DAInfer}},
we adopt \textsf{the gpt-3.5-turbo} model with the chat completions API to interpret the API semantic descriptions~\cite{GPT3.5}.
Specifically, we invoke the \textsf{ChatCompletion.create} interface to feed the constructed prompts to the LLM and fetch its response.
In our implementation, we set the temperatures for the two stages of prompting to 0.7.

To extract memory operation abstractions using embedding models, we employed the pre-trained models from the Sentence-BERT (SBERT)\cite{sentence-bert} and E5 \cite{wang2024multilinguale5textembeddings} frameworks. SBERT \cite{sentence-bert} is a modification of the BERT \cite{devlin-etal-2019-bert} architecture that uses Siamese and triplet network structures to generate semantically meaningful sentence embeddings. The fixed-size vectors in SBERT allow for highly efficient comparison. As a retrieval-first architecture, E5 \cite{wang2024multilinguale5textembeddings} excels at resolving asymmetric semantic similarity. This capability stems from its extensive pre-training on large-scale web corpora. We used these models to compute the embedding vectors for each simple sentence. Independently, we also computed the vectors for the four memory operation descriptions. The semantic correspondence between a sentence and each operation was quantified using cosine similarity. This approach allows us to align the linguistic functionality of the API methods with formal memory operations based on their shared semantic space.


\revision{We implement the neurosymbolic optimization based on Z3 solver~\cite{Z3, BjornerPF15}.
For any pair of APIs, we introduce $(n_1 + 1) \cdot (n_2 + 1)$ boolean variables to indicate whether the two API values are aliased,
where $n_1$ and $n_2$ are the numbers of the API parameters.
We directly encode the degree constraint and validity constraint symbolically,
while the semantic unit constraint and memory operation constraint are constructed and solved on demand,
relying on the outputs of our \textcolor{black}{desired NLP models}.
We count the number of boolean variables assigned to \textsf{True} and set them as the objective function.}
For better performance, we parallelize the invocations of the LLM in eight threads,
\deletion{To avoid redundantly applying the tagging model and the LLM, we}
\revision{and} introduce the memorization technique to store the tagging result and \textcolor{black}{the result of memory operation abstraction} upon each semantic description.
If a word or an API semantic description has been processed before, we directly reuse the previous result.

\section{Evaluation}
\label{sec:eval}

\revision{We evaluate \ToolName\ by investigating the following research questions:}
\deletion{To quantify the effectiveness of our approach, we propose the following research questions:} 

\begin{itemize}[leftmargin=*]
	\item \textbf{RQ1: } How accurately and efficiently does \ToolName\  generate data-flow and aliasing specifications?

	\item \textbf{RQ2: } How does \ToolName\ benefit library-aware static analysis clients?
      
    \item \textbf{RQ3: } How does \ToolName\ compare against other approaches?

\end{itemize}

\subsection{Experimental Setup}
All the experiments in \textsc{DAInfer} \cite{DAInfer} are performed on a 64-bit machine with 40 Intel(R) Xeon(R) CPU E5-2698 v4 @ 2.20 GHz and 512 GB of physical memory. For \ToolName, we used a machine with the same CPU setup, 251 GB of physical memory, and an \textsc{NVIDIA RTX 3090 GPU} for hosting LLMs and embedding models.
We invoke the Z3 SMT solver with its default options.

\smallskip
\emph{\textbf{Subjects.}}
To show the superiority of \ToolName\ \textcolor{black}{in producing alias specifications},
we evaluate \textsc{Atlas}~\cite{Bastani0AL18}, \textsc{USpec}~\cite{EberhardtSRV19}, and \ToolName\
upon the same set of Java classes.
Specifically, the Java classes are collected from:
(1) The classes of which the specifications are manually specified in \textsc{FlowDroid}~\cite{Arzt14FlowDroid};
(2) The classes appearing in the inference results of \textsc{USpec}~\cite{EberhardtSRV19}.
Since the dataset of \textsc{Atlas}~\cite{Bastani0AL18} is not publicly available, we cannot conduct experiments on it. 
In total, our benchmark contains 167 Java classes offering 8,342 APIs,
which range from general-purpose libraries, including Android framework and Java Collections Framework, to specific-usage libraries, such as \textsf{Gson}.
Without ambiguity, we call the first and the second kinds of classes from \textsc{FlowDroid} benchmark and \textsc{USpec} benchmark, respectively.

\begin{figure*}[t]
	\centering
	\includegraphics[width=\linewidth]{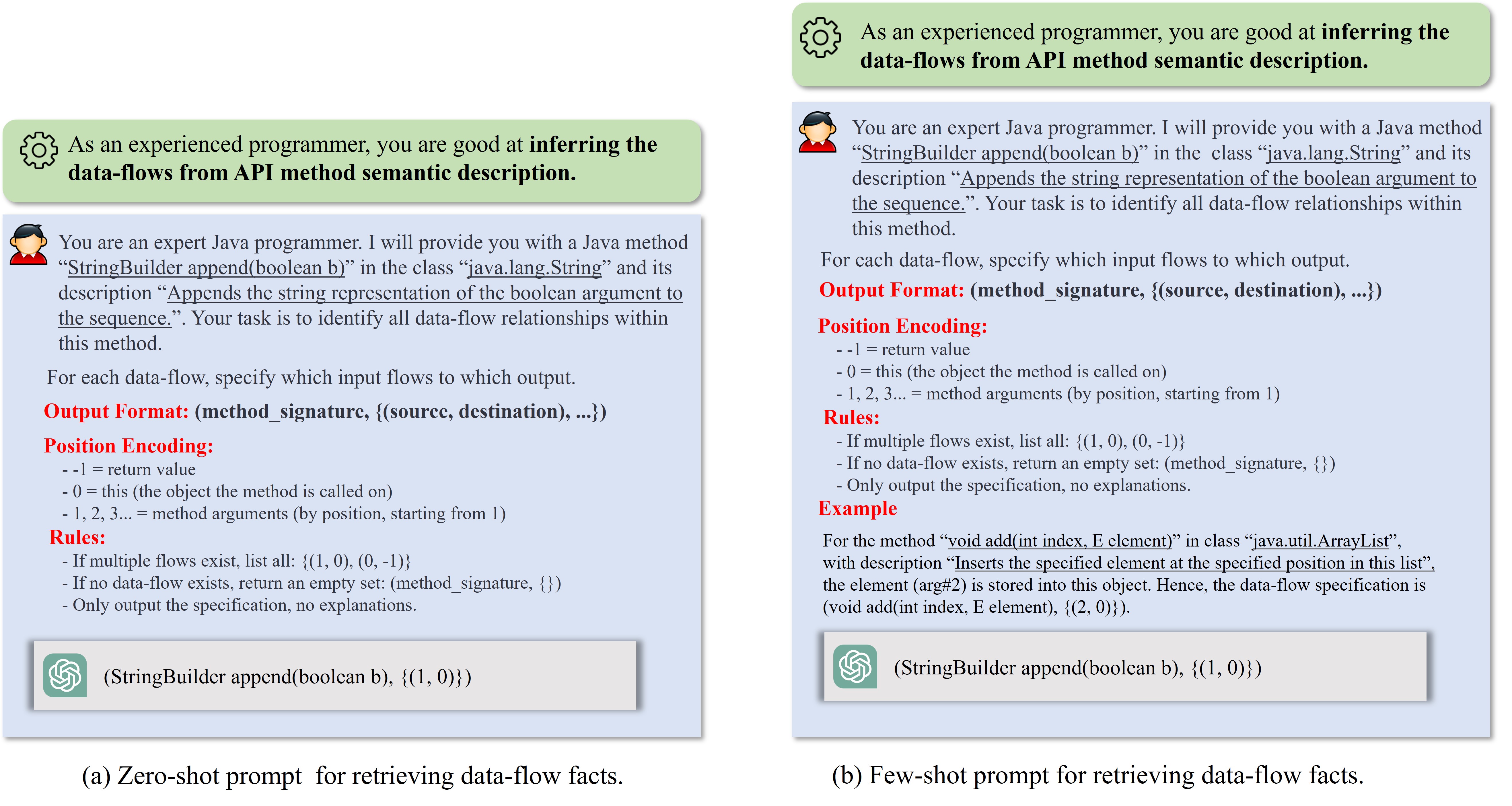}
	\vspace{-7mm}
	\caption{\revision{The zero and few-shot prompt templates for inferring data-flow specifications for an API method.}}
		\vspace{-3mm}
	\label{fig:prompt-dataflow}
\end{figure*}

\textcolor{black}{To evaluate the effectiveness of our memory operation abstraction module in inferring data-flow facts for constructing specifications, we evaluated \ToolName\ across a diverse suite of state-of-the-art LLMs trained on both general-purpose and programming-specific datasets. Our selection includes \textsf{deepseek-V2} \cite{deepseekai2024deepseekv2strongeconomicalefficient}, \textsf{gpt-4o-mini} \cite{openai2023gpt4}, \textsf{qwen2.5-coder} \cite{hui2024qwen25codertechnicalreport}, and \textsf{deepseek-coder-V2} \cite{deepseek-coder}. We employed the zero-shot and few-shot prompting strategies (illustrated in Figure \ref{fig:prompt-dataflow}) to task these models with inferring data-flow specifications. For each API method, the models were provided with the method signature, its containing class, and the corresponding documentation. To facilitate a rigorous comparison, we leveraged the same 167 Java classes originally specified in the FlowDroid framework, comparing the specifications generated by our tool against those generated by the LLMs. }

\subsection{Data-Flow Specification Inference}
\label{subsub:memoryoperation}
\ \ \ \  \emph{\textbf{Effectiveness.}}
To assess the effectiveness of our approach, we use the manually curated dataset of data-flow specifications from FlowDroid \cite{Arzt14FlowDroid} as our ground truth. Due to the extensive number of specifications in the original dataset, we randomly selected 141 classes comprising 1,094 API methods. Among these, 1,064 methods provide semantic descriptions within their documentation. After excluding 19 deprecated methods, our final evaluation set consisted of 1,045 methods. To ensure the accuracy of the ground truth and address potential flaws in the original collection, two researchers independently re-labeled the dataset by analyzing the source code and documentation of each API method to verify the corresponding data-flow specifications. 

As demonstrated in Table \ref{tab:results}, \ToolName\ achieves its maximum effectiveness when utilizing embedding models rather than LLMs. General-purpose LLMs, such as \textsf{gpt-4o-mini} and \textsf{deepseek-v2}, achieve relatively high precision but lower overall recall, even with few-shot prompting. This indicates a tendency toward under-approximation; these models are often too conservative, failing to identify valid data-flows described in the documentation. However, when they do predict flows, they are still prone to spurious over-approximation in specific cases. While few-shot prompting techniques improve their recall, these models remain prone to generating spurious flows. For instance, ``\textsf{gpt-4o-mini}'' infers a data-flow from the first parameter of the API method \textsf{android.os.Bundle.getStringArrayList(string)}  to ``\textsf{this}'' (the ``\textsf{Bundle}'' object). In addition, code-specialized LLMs achieve higher precision (\(>80\%\)), but they still struggle to reach high recall. For example, \textsf{qwen2.5-coder} incorrectly infers a data-flow from the first parameter of \textsf{android.content.Intent.getStringArrayExtra(string)} to ``\textsf{this}'' (the ``\textsf{Intent}'' object). The lower false positive rate in code-specialized models is due to their training on structural code patterns; however, they remain limited by under-approximation. By being overly conservative in predicting flows, these models frequently fail to identify valid data-flows, thereby increasing their false negative rates.

\begin{figure}[htbp]
    \centering
    \begin{subfigure}[t]{1.0\textwidth} 
        \centering
        \includegraphics[width=\textwidth]{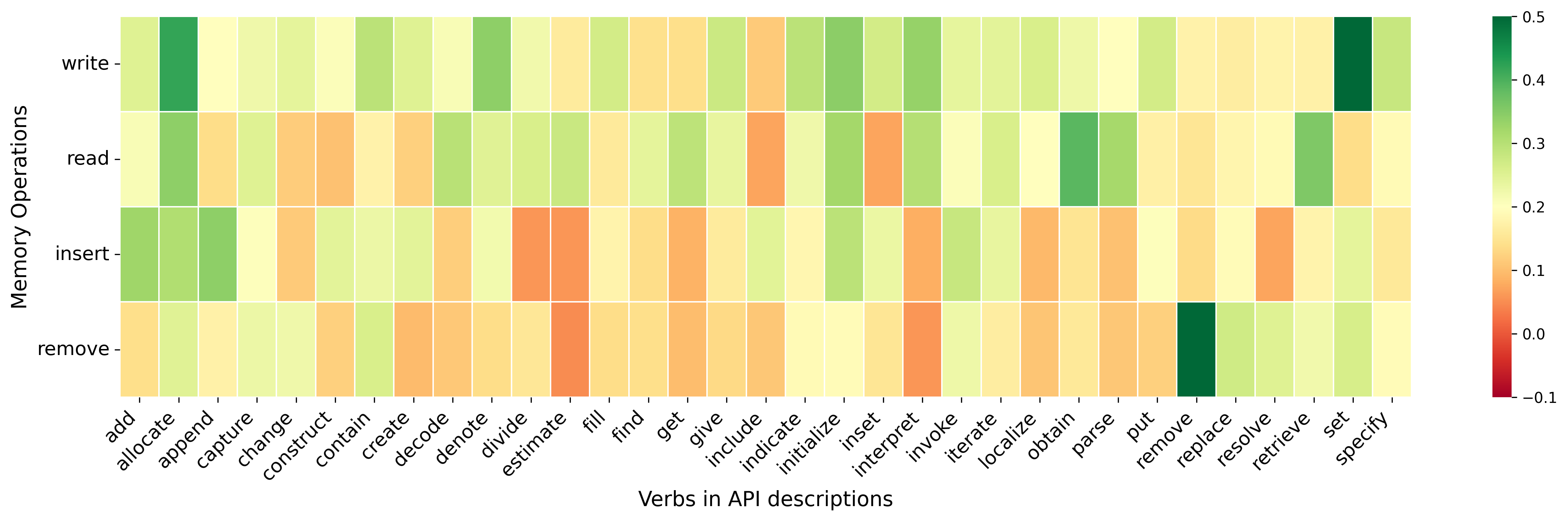}
        \caption{\textsc{SBERT-MPNet} embedding model.}
        \label{fig:heatmap_mpnet}
    \end{subfigure}
    
    \vspace{1em} 
    
    \begin{subfigure}[t]{1.0\textwidth} 
        \centering
        \includegraphics[width=\textwidth]{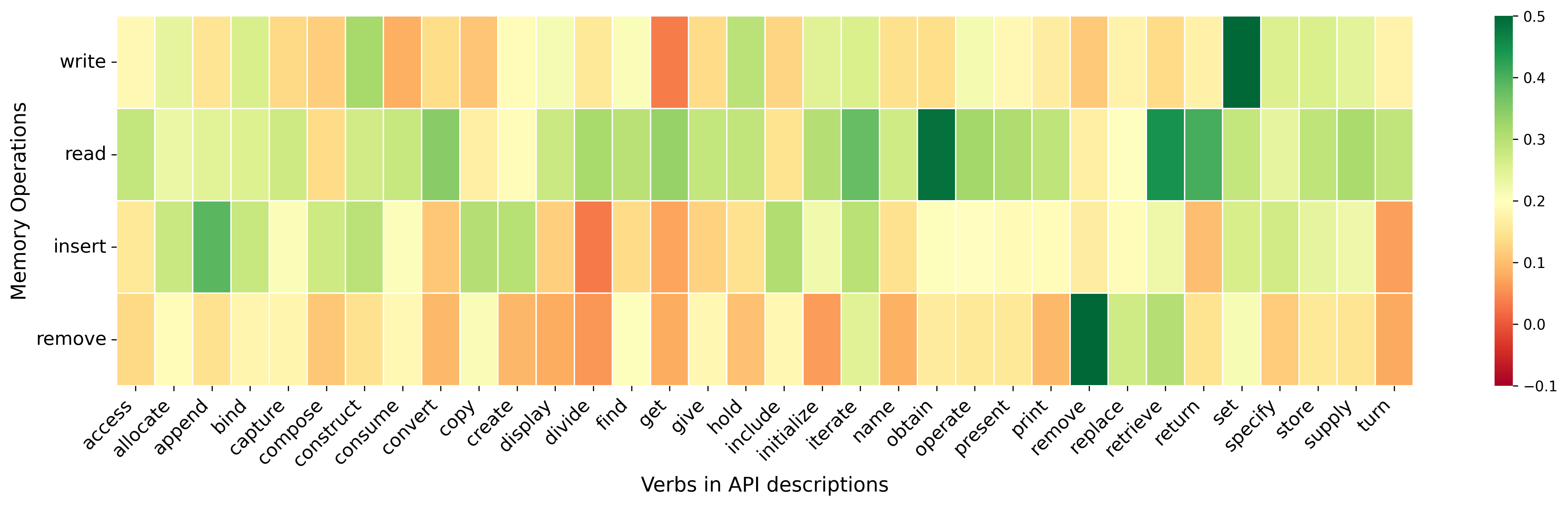}
        \caption{\textsc{SBERT-Mini} embedding model.}
        \label{fig:heatmap_mvm16}
    \end{subfigure}
    
    \caption{Cosine similarity scores for various verbs in API method descriptions with verbs used in our designated memory operation descriptions with different models of SBERT.}
    \label{fig:heatmaps_comparison}
\end{figure}

Embedding models demonstrate superior performance in both recall and precision compared to LLMs. As shown in Table \ref{tab:results}, these models consistently achieve higher recall and precision scores exceeding 82\% and 88\%, respectively, indicating higher reliability for specialized data-flow specification inference. This success is primarily due to the semantic alignment between the verbs used in API method descriptions and the definitions of memory operations. The heatmaps in Figure \ref{fig:heatmaps_comparison} prove this claim by illustrating the similarity scores between various verbs found in API documentation with our memory operation descriptions using two popular models of \textsc{SBERT}. Notably, these figures reveal that most descriptive verbs align closely with the correct memory operations.

\begin{table}[ht]
    \centering
	\vspace{-2mm}
    \caption{Accuracy (Acc.), Recall (Rec.), Precision (Pre.), and F1-score for data-flow specification retrieval.}
	\resizebox{0.65\linewidth}{!} { 
        \begin{tabular}{c|c|K|K|K|G}
            \hline
            \textbf{Type} & \makecell[c]{\textbf{Model}}  & \multicolumn{1}{c|}{\textbf{Acc.}} & \multicolumn{1}{c|}{\textbf{Rec.}} &  \multicolumn{1}{c|}{\textbf{Pre.}} & \multicolumn{1}{c}{\textbf{F1-score}} \\ 
            \hline            
            \multirow{5}{*}{\makecell{\textbf{zero-shot } \\ \textbf{prompting}}} & deepseek-v2 & 23.83 & 24.69 &  87.29 &  0.38 \\
            & gpt-4o-mini	 &   61.83 &  72.25 &  81.09 &  0.76 \\
            & qwen2.5-coder &  60.49 &  65.56 &  87.66 &  0.75 \\
            & deepseek-coder-v2 &  47.51 &  54.74 &  78.23 &  0.64 \\
            \hline 
            \multirow{5}{*}{\makecell{\textbf{few-shot } \\ \textbf{prompting}}} & deepseek-v2 &  49.26 &  64.95 &  67.09 &  0.66 \\
            & gpt-4o-mini	 &   69.30 &  76.46 &  87.09 &  0.81 \\
            & qwen2.5-coder &  72.05 &  75.08 &  94.68 &  0.83 \\
             & deepseek-coder-v2	&  50.74 &  58.05 &  80.11 &  0.67 \\

             \hline 
              \multirow{5}{*}{\makecell{\textbf{embedding} \\ \textbf{models}}} &  \makecell[l]{\textsc{SBERT-Mini} } &  74.93 &  82.36 &  87.25 &  0.86 \\
               &  \makecell[l]{\textsc{SBERT-MPNet} } &  75.6 &  83.9 &  89.42 &  0.86 \\
               &  \makecell[l]{\textsc{E5-Base}} &  73.32 &  83.95 &  85.37 &  0.85 \\
               &  \makecell[l]{\textsc{E5-Large}}&  76.34 &  84.69 &  87.5 &  0.87 \\
        \bottomrule
        \end{tabular}
    }
	\vspace{-3mm}
    \label{tab:results}
    
\end{table}

\emph{\textbf{Efficiency.}}
To evaluate the efficiency of embedding models relative to LLMs, we utilize two primary metrics: (1) inference time, and (2) cost. As indicated in Table \ref{tab:performance_comparison}, the selected LLMs require significantly longer durations to infer data-flow specifications compared to their embedding counterparts. Even in the best-case scenario, where \textsf{deepseek-coder-v2} achieves its peak performance, its analysis time remains orders of magnitude higher than that of the embedding models. Furthermore, the throughput of embedding models (e.g., \textsc{SBERT-MPNet} and \textsc{E5-base}) significantly exceeds that of LLMs like \textsf{qwen2.5-coder} in few-shot mode. Here, throughput is defined as the volume of data processed per second (measured in tokens for LLMs and vectors for embedding models). On average, the LLMs processed 8,389 prompts to achieve inference, whereas the embedding models extracted and encoded 1,147 simple sentences. This empirical evidence confirms that for high-volume data-flow inference, the ``\textsf{Encoder-only}'' architecture provides a superior efficiency-to-performance ratio compared to the ``\textsf{Decoder-only}'' autoregressive approach, which is fundamentally bottlenecked by sequential, token-by-token generation. 

To quantify the total cost, we monitored resource utilization across an array of \textsc{NVIDIA RTX 3090} GPUs. Our findings reveal a significant disparity in VRAM overhead. Even when utilizing 4-bit quantization to minimize memory requirements, the local LLM suite (\textsf{qwen2.5-coder-32B} and \textsf{deepSeek-coder-v2}) requires approximately 20-30 GB of VRAM just to host these models. Comparatively, the embedding models occupy at most 2GB of the memory. Furthermore, we analyzed the operational costs considering the rental of a cloud service with the same setup with our environment. In addition, \textsf{gpt-4o-mini} consumed 4M output and 73K input tokens for retrieving the data-flow specifications of the entire dataset, totaling \$2.41 USD. While this API-based approach is economically viable for one-off inferences, the \textbf{total time} required for LLM-based inference is several orders of magnitude higher than our embedding-based approach, which processes the entire dataset in less than a few seconds. Even when utilizing the same hardware environment, the embedding models (\textsf{E5}, \textsf{SBERT}) demonstrate a massive advantage in computational efficiency. 

\begin{table}[h]
\centering
\caption{Comparative analysis of resource intensity between embedding models and LLMs with few-shot prompting. }
\label{tab:performance_comparison}
\begin{tabular}{@{}c|c|c|c|c@{}}
\hline
\multirow{2}{*}{\textbf{Model}} & \textbf{Parameters} & \textbf{Throughput} & \textbf{Cost} & \textbf{Time Cost} \\ 
& & \textbf{(Items/sec.)}& \textbf{(per 1M items)} & \textbf{(sec.)}\\
    \hline
    \textsf{deepseek-v2} & 16B & 1.62 & \$0.11\textsuperscript{*}  & 3,358.2  \\
    \textsf{qwen2.5-coder}  & 32B & 2.26 &  \$0.08\textsuperscript{*} &  2,344 \\
    \textsf{gpt-4o-mini} & proprietary & 5.73 & $\$0.15-\$0.6$\textsuperscript{+}& 927\\
    \textsf{deepseek-coder-v2} & 7B & 2.26 &  \$0.03\textsuperscript{*}  & 900\\
    \textsc{SBERT-Mini} & 22M & 6966.66  & $\$~$0.0\textsuperscript{\textdagger}   &  0.15\\ 
    \textsc{SBERT-MPNet} & 110M & 2223.40 & \$~0.0\textsuperscript{\textdagger}   &  0.47\\ 
    \textsc{E5-base} & 110M & 460.35 & \$0.0\textsuperscript{\textdagger}  & 2.27 \\ 
    \textsc{E5-large} & 335M & 176.55 & \$0.0\textsuperscript{\textdagger}   &  5.19\\ 
\bottomrule
\noalign{\smallskip}
\multicolumn{5}{l}{\textsuperscript{*} \small Estimated based on a market rental rate of USD \$0.56/hr for equivalent hardware (RTX 3090, 215GB RAM). } \\
\multicolumn{5}{l}{\textsuperscript{\textdagger} \small Embedding models incur negligible costs on standard hardware and values are rounded to the nearest cent. }\\
\multicolumn{5}{l}{\textsuperscript{+} \small Based on OpenAI official API pricing for input/output token blends.}\\
\end{tabular}
\end{table}


\subsection{Alias Specification Inference }
\label{subsec:effective_efficient}
\ \ \ \  \emph{\textbf{Effectiveness.}}
\revision{
Although \textsc{USpec} offers the raw data and the source code of \textsc{Atlas} is available, the ground truth used in the two previous studies is not published. 
Also, the specifications offered by \textsc{FlowDroid} are manually specified by the developers, and thus, may contain several flaws and miss several correct ones.
Hence, we have to label the specifications of the benchmarks manually.
}
Meanwhile, investigating all the classes demands tremendous manual effort.
Following the recent study~\cite{EberhardtSRV19}, we randomly select 60 classes that offer 2,771 APIs in total.
For each API, we examine whether it forms store-load API pairs with other APIs offered by the same class,
of which the number can reach 50 on average.
\revision{
To make the manual examination more reliable, we invite five experienced engineers from the industry as volunteers to specify the specifications independently. Specifically, they refer to the specifications specified by the developers of \textsc{FlowDroid} and inferred by existing works (i.e., \textsc{USpec} and \textsc{Atlas}), and meanwhile, investigate the library documentation and implementation simultaneously. In the end, we merge the specifications specified by the five volunteers and resolve the inconsistent parts following the principle of max voting, eventually obtaining 988 API aliasing specifications as the ground truth.}

According to our investigation, we find that \ToolName\ achieves high precision and recall upon the experimental subjects.
In total, it successfully infers 2,680 API aliasing specifications.
For the randomly selected 60 classes, \textsc{DAInfer} \cite{DAInfer} infers 1,019 API aliasing specifications, 813 of which are correct,
achieving a precision of 79.78\%.
we discover that \textsc{DAInfer} misses specifications,
achieving a recall of 82.29\%.
\textcolor{black}{In our latest study, \ToolName\  achieves 79\% precision and 88\% recall when using \textsc{SBERT-MPNet}. By systematically decomposing complex descriptions and inferring abstract memory operations based on \textbf{sentence structure}, \ToolName\ effectively filters out non-functional linguistic noise. This structural approach significantly reduces false positives, as it isolates the primary action from the surrounding constraints and conditions that frequently mislead general-purpose LLMs. }

After examining all the APIs of the selected classes,
Interestingly, we collect the specifications where the API names contain \text{``}get\text{''} or \text{``}set\text{''},
and discover that such specifications only take up 33.49\% of all the inferred ones.
It shows that \textsc{DAInfer} can understand how APIs operate upon the memory even if diverse verbs are used.
We also compare our results with the specifications in the \textsc{FlowDroid} and \textsf{USpec} benchmarks. 
It is shown that \ToolName\ infers 170 out of the total 210 specifications in \textsc{FlowDroid} benchmark and 65 out of the total 82 specifications inferred by \textsc{USpec},
achieving 81.0\% and 79.3\% recall upon the two benchmarks, respectively.
The above results show that \textsc{DAInfer}, and \ToolName, the latest update of it with embedding models, can effectively infer the API aliasing specifications from documentation.

\begin{table}[t]
	\caption{Efficiency of \textsc{DAInfer} and its ablations. \textbf{\# Input} represents vectors for embedding models and prompts for LLMs.}
	\vspace{-2mm}
	\label{eval:efficiency}
	\resizebox{0.8\linewidth}{!} {
		\begin{tabular}{c|c|c|c|c}
			\hline
			\textbf{Tool} & \textbf{\# Tagging} & \textbf{\# Inputs} & \textbf{\# Tokens} & \textbf{Time Cost (sec)}  \\ \hline
            \ToolName\ (\textsc{SBERT-Mini}) & 32,325 & 3,597 & NA  & 87.86 \\
			\ToolName\ (\textsc{SBERT-MPNet}) & 32,325 & 3,597 & NA & 88.88 \\
			\textsc{DAInfer} \cite{DAInfer}   & 32,325  &   2,950     & 726,425  &  892.93      \\ 
			\textsc{DAInfer-Type} \cite{DAInfer}  &  32,325   &  5,164    &   1,276,254  &     1,734.63   \\ 
			\textsc{DAInfer-Exhaustive}\cite{DAInfer}   & 58,846  & 8,090   &  1,994,017   &  2,844.26   \\ \hline		
		\end{tabular}
	}
	\vspace{-3mm}
\end{table}

\smallskip
\emph{\textbf{Efficiency.}}
We quantify the efficiency of \textsc{DAInfer} and \ToolName\ with four metrics, including the number of times the tagging model is applied, the input count (representing LLM prompts or vector calculations),  the token cost, and the time cost.
As shown in Table~\ref{eval:efficiency}.
\textsc{DAInfer} \cite{DAInfer} applies the tagging model 32,325 times and interacts with the LLM 2,950 times using 726,425 tokens,
and the overall time cost is 892.93 seconds (around 15 minutes).
According to the OpenAI billing strategy, 
we only need to pay 1.09 USD in total.
We also conduct the ablation study to demonstrate the benefit of the neurosymbolic optimization algorithm.
Specifically, the ablation \textsc{DAInfer-Exhaustive} applies the two NLP models to all the APIs, while the ablation \textsc{DAInfer-Type} applies the NLP models to the APIs satisfying the degree constraint and the validity constraint.
As shown in Table~\ref{eval:efficiency}, 
\textsc{DAInfer-Type} invokes the LLM 5,164 times with 1,276,254 tokens in total and finishes analyzing all subjects in 1,734.63 seconds.
Besides, \textsc{DAInfer-Exhaustive} has to apply the tagging models 58,846 times and invoke the LLM 8,090 times using 1,994,017 tokens,
of which the whole process finishes in 2,844.26 seconds.
The key reason for the differences between the ablations is that the solving steps at \revision{Lines 4 and 7}\deletion{Lines 5 and 9} in Algorithm~\ref{alg:solving} can effectively reduce the number of times the tagging model and the LLM are applied, respectively.
Compared to \textsc{DAInfer-Type} and \textsc{DAInfer-Exhaustive},
\textsc{DAInfer} achieves the inference with $1.94\times$ and $3.19\times$ speed-ups when relying on two-staged prompting with LLMs. 

By utilizing \ToolName\ with embedding models and an optimized tagging-based inference approach, the specification retrieval process is significantly accelerated. From the original data, we extract a total of 3,597 simple sentences averaging 87.4 seconds. The subsequent memory operation abstraction using embedding models requires at most 2 seconds, specifically 0.48 seconds for \textsc{SBERT-Mini} and 1.48 seconds for \textsc{SBERT-MPNet}. Consequently, the overall speed-up is enhanced by $10.16\times$ and $10.04\times$ for \textsc{SBERT-Mini} and \textsc{SBERT-MPNet} respectively, compared to using \textsc{DAInfer} with LLMs.  
Hence, our neurosymbolic optimization can efficiently support the specification inference.

\subsection{Effects on Client Analysis}
\label{sec:ec}
Following existing studies~\cite{Bastani0AL18, EberhardtSRV19}, we choose alias analysis and taint analysis as two fundamental clients of \ToolName\ to quantify its effects.

\begin{wrapfigure}[11]{r}{0.45\textwidth}
\vspace{-10mm}
	\centering
	\scalebox{1.0}{
		\begin{minipage}{0.45\textwidth}
			\centering
			\begin{figure}[H]
				\centering
				\includegraphics[width=\linewidth]{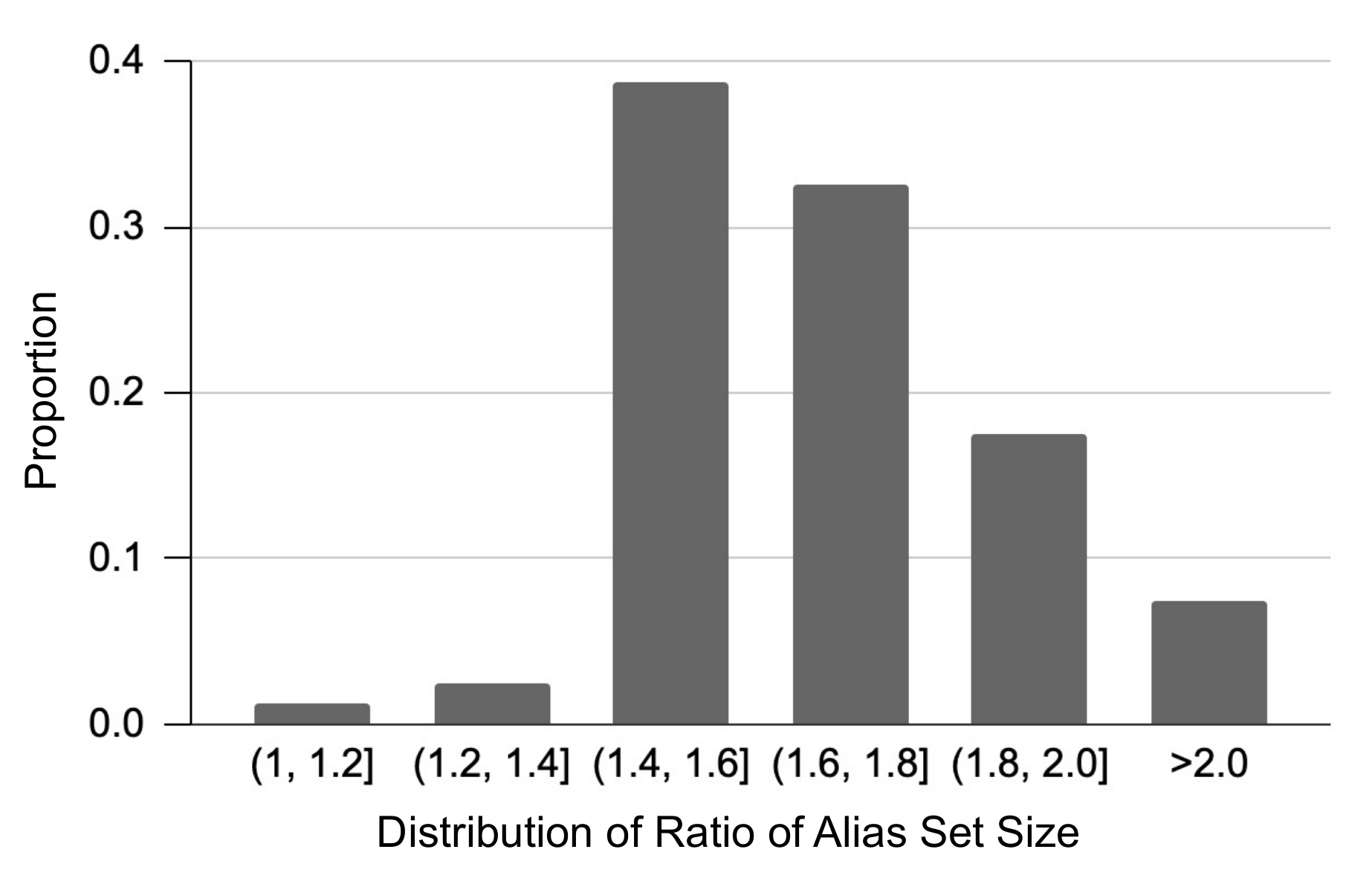}
				\vspace{-8mm}
				\caption{The results of pointer analysis}
				\label{fig:client}
				\vspace{-8mm}
			\end{figure}
		\end{minipage}
	}
\end{wrapfigure}

\emph{\textbf{Effect on Alias Analysis.}}
We conduct the field and context-sensitive alias analysis by running a static analyzer \textsc{Pinpoint}~\cite{Shi18Pinpoint, Falcon} upon 15 Java projects in two settings.
In the setting \textsf{Alias-Empty},
we provide empty specifications of library APIs, i.e., discarding all the possible alias facts introduced by library API calls.
In the setting \textsf{Alias-Infer},
we apply the inferred correct API aliasing specifications to the pointer analysis.
\revision{For each given pointer, \textsc{Pinpoint} computes its alias facts in a sound manner.}
We quantify the alias set sizes of the return values of library APIs
\revision{
and compute $\frac{size_{\textsf{infer}}}{size_{\textsf{empty}}}$ for each library API invocation, 
where $\textsf{size}_{\textsf{infer}}$ and $\textsf{size}_{\textsf{empty}}$ are the alias set sizes of the return value under the settings \textsf{Alias-Infer} and \textsf{Alias-Empty}, respectively.
Figure~\ref{fig:client} is the histogram showing the distribution of the ratios of alias set sizes.
According to the ratios of alias set sizes, we can discover that the average increase ratio reaches 80.05\% with the benefit of our inferred specifications. Except for the intervals (1, 1.2] and (1.2, 1.4], the size increase ratio is larger than 40\% as the ratio is larger than 1.4. The proportion of such library API invocations reaches 96.25\%.}
\deletion{
On average, the alias set sizes of the return values are increased by 80.05\%
with the benefit of our inferred specifications.
The alias set sizes of 96.25\% return values of the library API calls are increased by at least 40\%.}
Because our pointer analysis is sound \revision{and we investigate the same set of return values of library API calls}, the increases in the alias set sizes demonstrate that \ToolName\ promotes the alias analysis in discovering more alias facts in the applications using libraries.

\smallskip
\emph{\textbf{Effect on Taint Analysis.}}
We choose three different settings of specifications for \textsc{FlowDroid} to conduct the taint analysis, namely \textsf{Taint-Empty}, \textsf{Taint-Manual}, and \textsf{Taint-Infer}.
Here, \textsf{Taint-Empty} and \textsf{Taint-Infer} are similar to the two settings in the pointer analysis,
\revision{and the sources and sinks are specified based on the default taint specification offered by \textsc{FlowDroid}.}
Under the setting \textsf{Taint-Manual}, we apply the manual specifications provided by \textsc{FlowDroid} directly.
We select 23 popular Android applications in F-Droid~\cite{FDroid},
which cover different program domains,
including navigation, security, and messaging applications.

\begin{wrapfigure}[11]{r}{0.46\textwidth}
	\vspace{-9mm}
	\centering
	\scalebox{1.0}{
		\begin{minipage}{0.45\textwidth}
			\centering
			\begin{figure}[H]
				\centering
				\includegraphics[width=\linewidth]{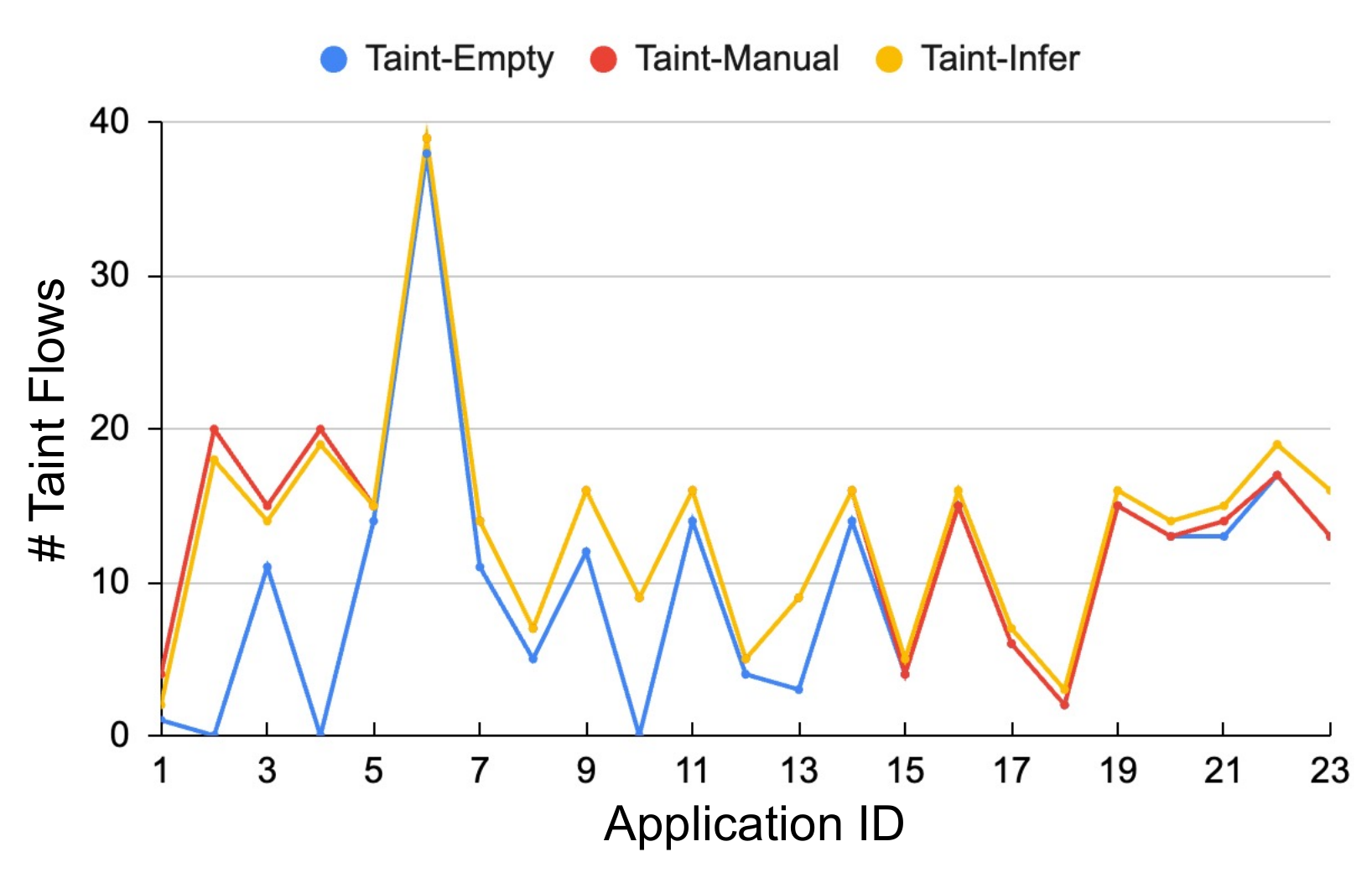}
				\vspace{-8mm}
				\caption{The results of taint analysis}
				\label{fig:taintclient}
				\vspace{-4mm}
			\end{figure}
		\end{minipage}
	}
\end{wrapfigure}
\smallskip
Figure~\ref{fig:taintclient} shows the number of taint flows discovered under the three settings.
Specifically, \textsc{FlowDroid} discovers 225 taint flows under \textsf{Taint-Empty}, while it finds 304 taint flows under \textsf{Taint-Manual}.
Notably, 79 out of 304 taint flows are induced by the aliasing relations among API parameters and returns. 
When we run \textsc{FlowDroid} under \textsf{Taint-Infer}, it discovers 310 taint flows, 85 of which are discovered based on the correct API aliasing specifications inferred by \ToolName.
There are six taint flows in three apps not discovered by \textsc{FlowDroid} under the setting \textsf{Taint-Infer} due to false negatives of our inference algorithm.
However, 12 taint flows discovered under \textsf{Taint-Infer} are not discovered under \textsf{Taint-Manual}.
The results demonstrate that \ToolName\ promotes the taint analysis in discovering more taint flows.
\revision{We do not seek the confirmations of taint flows,
which may depend on the developers' subjective intentions and the choices of taint specifications.
However, the ability to discover more taint flows has shown the practical impact of \ToolName\ in detecting potential taint-style vulnerabilities.
This evaluation principle is also applied in many existing studies~\cite{Bastani0AL18, EberhardtSRV19, DBLP:journals/darts/SpathDAB16}.}

\subsection{Comparison with Existing Techniques}
\label{subsec:eval_comparison}
We initially compare \ToolName\ with \textsc{StubDroid} \cite{ArztB16} on constructing the data-flow specifications. 
Next, we compare \ToolName\ with the two most recent studies on API aliasing specification inference,
i.e., \textsc{Atlas}~\cite{Bastani0AL18} and \textsc{USpec}~\cite{EberhardtSRV19}.
Besides, we construct another baseline, \textsc{LLM-Alias},
which feeds the documentation to \textsc{ChatGPT} and generates API aliasing specifications via in-context learning.

\smallskip
\emph{\textbf{Comparison with StubDroid.}}
To evaluate the semantic richness of \ToolName, we compare the automated generation of data-flow specifications of \textsc{StubDroid}~\cite{ArztB16} with our solution. While \textsc{StubDroid} produces generic, method-level summary rules, \ToolName\ infers high-level memory primitives—specifically \textit{write}, \textit{read}, \textit{insert}, and \textit{remove}—for each API method.
By translating these primitives into data-flow \textit{gen} and \textit{kill} operations, \ToolName\ provides the semantic context necessary for precision in collection-heavy applications. 
Specifically, while \textsc{StubDroid} often leads to over-approximation by failing to identify when a taint should be invalidated, \ToolName\ utilizes the \textit{remove} primitive to trigger \textit{strong updates} (kills). 
This allows the engine to remove specific data-flow facts from the analysis state, thereby significantly reducing false positives.  
In addition, \textsc{StubDroid} fails to extract the data-flow specifications for abstract interfaces such as \textsf{java.util.List}. Hence, our assessments reveal that \textsc{StubDroid} achieves 46\% recall and 51\% precision in extracting the data-flow specifications for the selected classes we used earlier in section \ref{subsub:memoryoperation}. Comparatively, \ToolName\ with \textsc{SBERT-MPNet} achieves the recall of 82\% and the precision of 88\%. 
Furthermore, \textsc{StubDroid} require an average of 394.84 seconds to generate a specification for a single class (ranging from minimum of 0.36 to maximum of 10,242.88 seconds). Conversely, \ToolName\ infers these specifications from documentation in just a few seconds, as shown in Table \ref{tab:performance_comparison}
Unlike \textsc{StubDroid}, which relies on heavyweight bytecode analysis to discover flows, \ToolName\ infers memory operation abstractions, and then translates them directly into \textsf{FlowDroid}-compatible XML summaries in an efficient manner.

\smallskip
\emph{\textbf{Comparison with \textsc{Atlas}.}}
We run \textsc{Atlas}~\cite{AtlasRepo} upon the total 167 classes
and finish the inference in 74.48 minutes.
\revision{Note that the output of \textsc{Atlas} is the library implementation derived from unit test executions.
Automatically converting it into the specifications defined in Definition~\ref{def:spec} requires static analysis techniques.
Hence, we analyze the library implementation generated by \textsc{Atlas} with a field-sensitive pointer analysis,
which matches the store-load operations upon the same fields,
and eventually convert the output of \textsc{Atlas} to the API aliasing specifications defined in Definition~\ref{def:spec}.
For the classes labeled with ground truth in~\cref{subsec:effective_efficient},
\textsc{Atlas} infers 546 specifications and 454 correct ones, achieving 83.15\% precision and 45.95\% recall.
After investigating the results, we find that \textsc{Atlas} fails to generate the specifications for 111 classes in the experimental subjects, such as \textsf{android.os.Intent} and \textsf{android.os.Configuration}.
The root cause is that \textsc{Atlas} fails to infer the specifications
when the creation of library function parameters is non-trivial, or the unit test execution demands a specific environment,
such as an Android emulator.
In contrast, \ToolName\ can derive the API aliasing specifications for such classes.
}
Also, the aliasing specifications generated by \ToolName\ only depict the potential aliasing relations between parameters and return values,
while they all miss the preconditions under which such aliasing relations hold.
For example,
\textsc{Atlas} only obtains that the return value of \textsf{HashMap.get} can be aliased with the second parameter of \textsf{HashMap.put}, missing the precondition over their first parameter.
The restrictive templates used in the inference introduce the imprecision,
which is also reported in the prior study~\cite{EberhardtSRV19}. 
\deletion{
Lastly, it is worth noting that the output of \textsc{Atlas} is the library implementation derived from the unit test executions.
Automatically converting it into the specifications defined in Definition 2 requires static analysis techniques,
while the analyzers with different precision would yield different API aliasing specifications.
Hence, we do not quantify the precision and recall of \textsc{Atlas}'s inference results in a more fine-grained manner.}

\smallskip
\emph{\textbf{Comparison with \textsc{USpec}.}}
\textsc{USpec} is not open-sourced
due to its commercial use~\cite{EberhardtSRV19}.
To make the comparison, we asked the authors for the raw data of their evaluation.
According to their results,
\textsc{USpecs} successfully obtains 124 API aliasing specifications upon 62 classes.
Unfortunately, the precision of \textsc{USpec} only reaches 66.1\% (82/124).
\revision{For instance, \textsc{USpec} generates the incorrect aliasing specification $(\textsf{HashMap.put}, \textsf{HashMap.get}, \{ (0, 1) \}, 0)$ for the class \textsf{java.util.HashMap}.
The root cause is that \textsc{USpec} infers possible aliasing relations according to the usage events, 
while the keys and values of \textsf{HashMap} objects may have the same type, making the inference algorithm unable to distinguish them.
However, \ToolName\ successfully infers the specification via neurosymbolic optimization.}
We also quantify \textsc{USpec}'s recall based on our labeled specifications in~\cref{subsec:effective_efficient}.
It is shown that \textsc{USpec} misses 370 API aliasing specifications. The recall of inferring API aliasing specifications is only 18.14\%.
The root cause of its low recall is that \textsc{USpec} can only generate the aliasing specifications for the APIs used in the applications' code.

\smallskip
\emph{\textbf{Comparison with \textsc{LLM-Alias}.}}
We compare \ToolName\ with \textsc{LLM-Alias},
which directly queries \textsc{ChatGPT} with the documentation.
The response \textsc{ChatGPT} generates is a natural language sentence with an API aliasing specification.
Due to laborious effort,
we only examine the inference results for 60 classes that we randomly selected in~\cref{subsec:effective_efficient}.
The results show that \textsc{LLM-Alias} generates 801 API aliasing specifications for examined classes,
only 113 of which are correct,
yielding a precision of 14.11\% and a recall of 11.44\%.
Among 688 incorrect specifications, 60 specifications indicate the correct aliasing relations between parameters and return values,
while they do not pose any restrictions on API parameters as the preconditions.
The results show that vanilla LLMs without special designs have poor performance in understanding the concept of the aliasing relation.
In contrast, \ToolName\ achieves quite satisfactory precision and recall,
which benefits from our insightful problem reduction and efficient neurosymbolic optimization.

\begin{wrapfigure}[12]{r}{0.52\textwidth}
	\vspace{-6mm}
	\centering
	\scalebox{1.0}{
		\begin{minipage}{0.5\textwidth}
			\centering
			\begin{figure}[H]
				\centering
				\includegraphics[width=\linewidth]{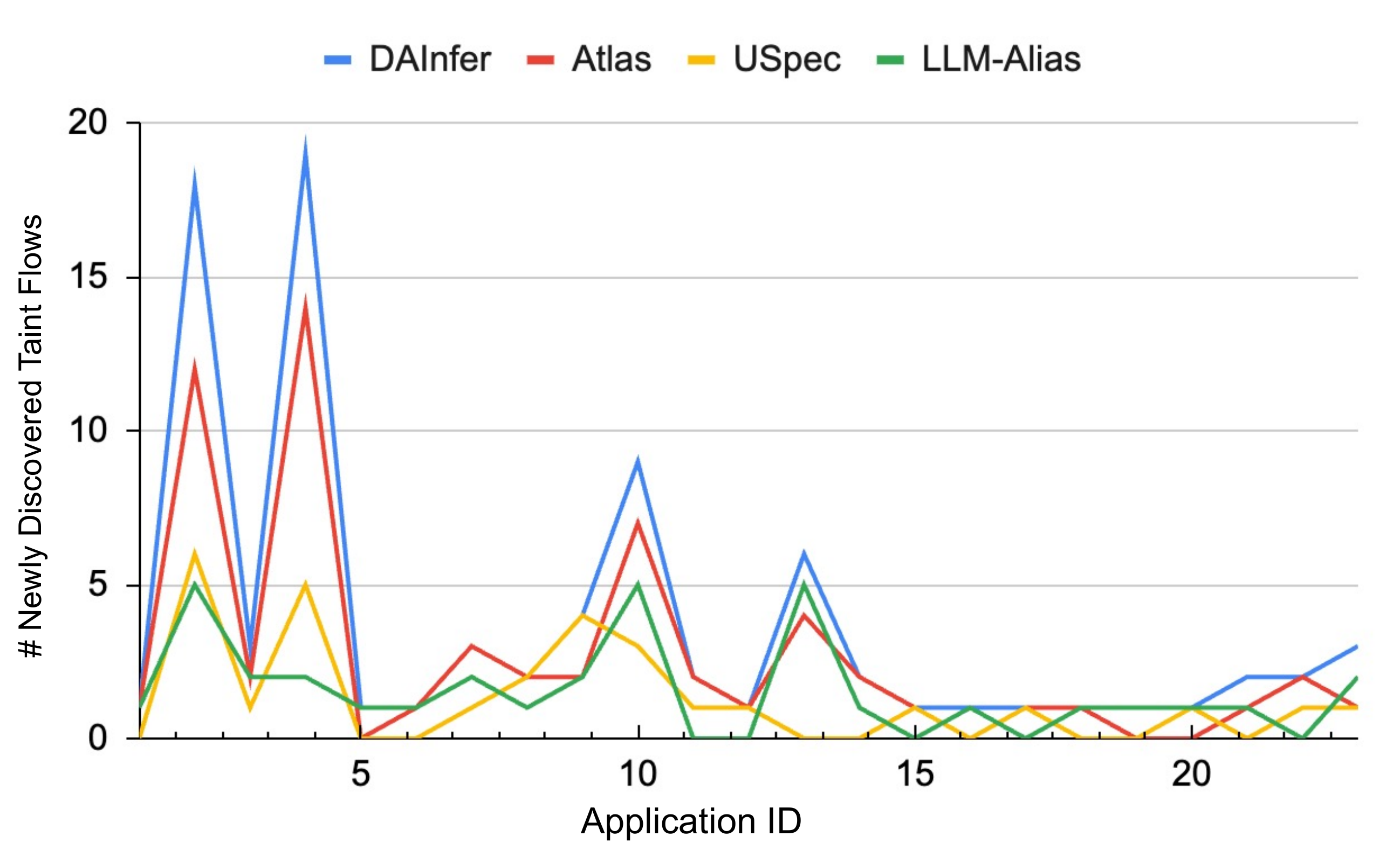}
				\vspace{-9mm}
				\caption{\revision{The results of taint analysis assisted with \textsc{Atlas}, \textsc{USpec}, and \textsc{LLM-Alias}}}
				\label{fig:taintclient_baseline}
			\end{figure}
		\end{minipage}
	}
\end{wrapfigure}

\smallskip
\textbf{Comparison upon Client Analyses.}
We also compare the effects of baselines on client analyses
with the same settings as the ones in~\cref{sec:ec}.
Specifically, \textsc{Atlas} introduces a 43.26\% increase in the alias set sizes on average,
which is lower than the one introduced by \ToolName.
\textsc{USpec} and \textsc{LLM-Alias} introduce 14.52\% and 12.17\% increase in the alias set sizes on average, respectively. Although \textsc{LLM-Alias} infers slightly more API aliasing specifications than \textsc{USpec},
the specifications inferred by \textsc{USpec} contribute more to the aliasing facts,
which might be caused by more frequent usage of the involved library APIs in the application code. 
\ToolName\ can introduce the highest average increase ratio in alias sets among different approaches.
Similarly, we find that \textsc{Atlas}, \textsc{USpec}, and \textsc{LLM-Alias} discover fewer taint flows than \textsc{DAInfer},
which is shown by Figure~\ref{fig:taintclient_baseline}.
Specifically, \textsc{DAInfer} newly discovers 85 taint flows,
while \textsc{Atlas}, \textsc{USpec}, and \textsc{LLM-Alias} detect 60, 29, and 35 taint flows in total, respectively.
Therefore, \ToolName\ has overwhelming superiority over existing techniques in assisting client analyses, including alias analysis and taint analysis.

\subsection{Limitations and Future Work}
\label{subsec:discussion}
Our approach has several drawbacks that demand further improvements.
\revision{First, \ToolName\ can not determine whether an API creates a new object.
When the developers create any new objects,
our inferred specifications can only depict data-flow facts instead of aliasing relations.
For example, \ToolName\ infers an API aliasing specification for \textsf{java.util.Map} that the return value of \textsf{Map.computeIfPresent} can be aliased with the second parameter of \textsf{Map.put} when their first parameters are aliased.
This is a wrong specification
as \textsf{computeIfPresent} returns \textsf{null} value or a newly computed value
instead of any existing values stored in the fields.}
Second, the semantic unit consistency requires two strings to be equal.
\revision{In our evaluation, however, we notice that several semantic units are not the same strings while they indicate the same concept in several rare cases.
For example, the first parameters of \textsf{SparseArray.set} and \textsf{SparseArray.valueAt} in the class \textsf{android.util.SparseArray} are \textsf{key} and \textsf{index}, respectively.
The two different strings are actually the indicators of the same semantic concept.
Hence, \ToolName\ can not infer the correct specification for the two APIs.
Although there are several traditional ways to extract synonyms for natural languages, such as WordNet~\cite{10.1145/219717.219748} and word embedding~\cite{zamani2017relevance},
they may fail to identify similar semantic units in programming languages, for example,
the similarity between \textsf{key} and \textsf{index} is measured to be even smaller than 0.1 by WordNet.
Even if we utilize several code models, such as \textsc{code2vec}~\cite{10.1145/3290353} and \textsc{CodeBERT}~\cite{feng-etal-2020-codebert}, 
they can still lead a false negative/positive when the similarity of the names in a correct/wrong specification is below/above the preset threshold.
}

\textcolor{black}{Furthermore, while relying on embedding models to infer memory operations and data-flow specifications enhances efficiency, these models are susceptible to false positives when API documentation is incomplete. For instance, developers often use cross-references—such as "\textsf{Please use writeBundle(Bundle) instead}''"—rather than providing redundant descriptions for similar methods or omitting return descriptions for state-updating API methods like \textsf{java.io.IntBuffer.put(string)}. To mitigate this, we refined the description analysis by manually augmenting missing contextual information. Additionally, certain verbs such as ``\textsf{copy}'' initially yielded low similarity scores for their corresponding memory operations. In addition, the dependency parser occasionally struggled with complex structures. Specifically, it erroneously classified ``\textsf{Set the point’s x and also the sentence y coordinates}'' as a compound sentence, leading to misleading results.
}

To further improve \ToolName, we can explore several directions in the future.
\revision{First, we can leverage domain-specific LLMs or \textcolor{black}{embedding models} for code, which allow local deployment, to validate the semantic unit consistency. If the inference of general-purpose LLMs, such as GPT-4, becomes much more efficient and cheaper in the future, we can also prompt them directly without introducing significant overhead.
The above models can hopefully support us in identifying the semantic units indicating the same concept, even if they are not the same string.}
\revision{Second, \ToolName\ requires a manually specified parser for documentation. Since LLM latency remains a factor, developing more robust dependency parsers tailored for technical documentation will be critical for accurate verb identification and relationship mapping.}
\textcolor{black}{Third, domain-specific fine-tuning of the embedding models would enhance the identification of memory-manipulation verbs, bridging the semantic gap between general-purpose language and API specifications. These general models are trained on broad corpora like \textsc{Wikipedia}, causing a semantic gap when interpreting specialized API documentation. For instance, a general model may not distinguish the subtle operational difference between ``\textsf{Transmitting an array}'' and `\textsf{Copying an array}'' yet these imply fundamentally different data-flow facts. For instance, pre-trained models like \textsc{CodeBERT} \cite{feng-etal-2020-codebert} offer bimodal understanding of code and text; however, they often struggle to distinguish specific memory actions in a zero-shot setting. Targeted training on technical corpora would allow embedding models to capture the precise memory semantics that general-purpose embeddings currently fail to resolve. }
\vspace{-1mm}
\section{Related Work}
\label{sec:related}

\ \ \ \ \emph{\textbf{Library Specification Inference.}}
The inference of library function specifications has always been a central topic in program analysis. 
Typically, IFDS/IDE-based approaches summarize the data-flow facts of libraries
as their semantic abstractions~\cite{rountev2008ide, ArztB16},
which can be reused across various clients of data-flow analysis. 
Established upon a symbolic memory model, 
shape analysis computes the memory state for each statement of a library function as invariants, 
and derives the preconditions/postconditions of each library function as its specification~\cite{Thomas02Parametric, Thomas10RelationalInterprocedural, Yang11BiAbduction}.
While the inferred specification accurately depicts the semantics of the library function, the analysis suffers from scalability problems, especially in the presence of complex program structures~\cite{ShapeSurvey20}. 

\textcolor{black}{To mitigate these limitations, }mining-based approaches leverage the program facts derived from applications to infer specific forms of specifications,
e.g., points-to~\cite{Bastani0AL18},
aliasing~\cite{EberhardtSRV19}, taint~\cite{ChibotaruBRV19},
and commutativity specifications~\cite{GehrDV15},
which support specific static analysis clients,
e.g., taint analysis~\cite{Arzt14FlowDroid} and Andersen-style pointer analysis~\cite{Pratik20SemanticModel}.
\revision{Another mining-based approach \textsc{AutoISES} automatically infers security specifications from high-quality application code and then guides the detection of security policy violations~\cite{DBLP:conf/uss/TanZMXZ08}.}
\textcolor{black}{Recent advancements, such as the \textsc{CSS framework} \cite{crosslanguagetaintanalysi}, extend these capabilities by generating caller-sensitive specifications for native code via iterative static analysis. Other efforts, including ModelGen \cite{modelgen} and Spectre \cite{spectre}, utilize dynamic analysis to identify data-flow and alias specifications at runtime; however, these techniques remain inherently constrained by input dependency and incomplete code coverage.}
Our work concentrates on the data-flow and aliasing specification inference,
which shares the same motivation as the existing studies~\cite{Bastani0AL18, EberhardtSRV19}
\textcolor{black}{while introducing a novel paradigm. Rather than relying on elusive code artifacts or limited execution traces, we leverage natural language documentation to unlock broader applicability. By employing embedding models, our method captures the latent data-flow intent within API descriptions. By synthesizing these semantic insights with named entity and type information, our framework employs optimization techniques to retrieve precise alias specifications independently of code analysis, thereby bypassing the visibility and scalability issues that hinder the state-of-the-art static and dynamic analyzers.}


\emph{\textbf{Natural Language Specification Understanding.}}
Natural language specifications, such as comments and documentation, are widely utilized in various software engineering tasks, including test case generation~\cite{DBLP:conf/issta/BlasiGKGEPC18, DBLP:conf/icse/MotwaniB19, DBLP:conf/sigsoft/ZhaiSPZLFM0020}, bug detection~\cite{DBLP:conf/kbse/ZhongZXM09, DBLP:conf/kbse/RenYXXXZS20, DBLP:conf/sigsoft/ZhaiSPZLFM0020, docflow}, and code search~\cite{DBLP:conf/icse/PanditaXZXOP12, prema}. Typically, \textsc{C2S}~\cite{DBLP:conf/sigsoft/ZhaiSPZLFM0020} employs semantic parsing to derive formal specifications from comments, which aids in test case generation and taint bug detection. Similarly, \textsc{Jdoctor}~\cite{DBLP:conf/issta/BlasiGKGEPC18} and \textsc{Swami}~\cite{DBLP:conf/icse/MotwaniB19} translate natural language specifications to formal ones to facilitate the generation of test cases covering exceptional behavior and boundary conditions, while they only focus on specific patterns, such as exceptions and numeric relations.

\textcolor{black}{In the realm of specification mining, SuSi \cite{susi} introduced supervised machine learning to classify sources and sinks for the information flow analysis. However, it relies heavily on manually engineered features, including both syntactic and semantic patterns in the API methods and their descriptions. \textsc{Doc2Spec} utilizes keywords, such as nouns and verbs indicating resource names and actions, respectively, to infer the resource specifications, which promote the resource misuse detection~\cite{DBLP:conf/kbse/ZhongZXM09}.}
\textcolor{black}{PreMA \cite{prema} extends this context to verb phrases to enhance the precision of detection of similar APIs. More closely related to our work, DocFlow \cite{docflow} uses contrastive learning to map resource names to sensitive categories (sources or sinks), and Fluffy \cite{fluffy} leverages a pre-trained embedding model (VarCLR \cite{vaclr}) to validate the taint flows based on API naming conventions. Despite this progress, these solutions typically require intensive supervised training and suffer from high false-positive rates in the absence of large, annotated datasets \cite{evaldocflow}.}

Although \ToolName\ shares similarities with existing works~\cite{DBLP:conf/kbse/ZhongZXM09, DBLP:conf/icse/PanditaXZXOP12} in terms of technical choices, such as named-entity recognition~\cite{DBLP:conf/muc/Chinchor98b}, our effort explores a new paradigm of deriving data-flow facts and aliasing relations from documentation, which can be generalized for other static analysis problems. 
A key innovation of our framework is employing a zero-shot embedding model to infer data-flow relations. Unlike traditional specifications mining solutions \cite{susi,docflow,fluffy}, our approach eliminates the need for large-scale manual labeling, offering a lightweight, scalable solution that remains effective in dynamic environments. Furthermore, by utilizing specialized embedding models rather than generative LLMs, we reduce the computational overhead while maintaining robust semantic inference.

\emph{\textbf{Large Language Models.}}
The Large Language Models (LLMs)~\cite{chatgpt,openai2023gpt4}, based on the decoder-only transformer architecture~\cite{DBLP:conf/nips/VaswaniSPUJGKP17}, are typically pre-trained on massive text corpora containing trillions of tokens. 
They exhibit exceptional zero/few-shot performance in a wide range of highly specific downstream tasks,
including
complex text generation~\cite{DBLP:journals/corr/abs-2107-03374},
interactive decision making/planning~\cite{DBLP:conf/iclr/ZhouSHWS0SCBLC23,DBLP:journals/corr/abs-2305-10601}, and tool utilization~\cite{DBLP:conf/iclr/YaoZYDSN023}.
Among various downstream tasks, 
reasoning task has traditionally been regarded as a typical challenge for LLMs~\cite{DBLP:journals/corr/abs-2110-14168},
which has attracted significant research interests.
Specifically, there has been a line of literature exploring the use of LLMs in automated theorem proving within formal logic. 
Pioneering studies~\cite{DBLP:conf/iclr/JiangWZL0LJLW23,DBLP:conf/nips/WuJLRSJS22,DBLP:conf/nips/JiangLTCOMWJ22} have focused on employing LLMs to generate proofs for theorems expressed in formal logic. 
Several recent efforts aimed to integrate advanced LLMs that have demonstrated impressive zero/few-shot performance in code completion tasks into formal logic reasoning tasks~\cite{DBLP:conf/nips/LampleLLRHLEM22,DBLP:conf/nips/WelleckLLHC22}. 
Inspired by these advancements, our previous research \cite{DAInfer} leveraged LLMs to interpret memory operation kinds, a sub-problem addressed within our current approach. 

Modern LLMs, particularly those fine-tuned on code such as \textsc{DeepSeek-Coder} \cite{deepseek-coder} and \textsc{Qwen2.5-Coder} \cite{hui2024qwen25codertechnicalreport}, have shown impressive zero-shot reasoning in code completion and formal logic tasks. However, for rigorous static analysis, directly applying generative LLMs presents significant risks of hallucination and computational latency at scale.  While LLMs offer powerful reasoning, our framework strategically balances their use with embedding models to optimize for both analytical depth and practical efficiency.

\section{Conclusion}
\label{sec:conclusion}

We proposed a new approach \ToolName\ to infer API aliasing specifications from documentation.
\ToolName\ adopts the tagging and NLP models to interpret informal semantic information in documentation.
It reduces the inference problem to an optimization problem that can be efficiently solved by our neurosymbolic optimization algorithm.
The inferred specifications are further fed to static analysis clients
for analyzing the applications using libraries.
\textcolor{black}{Our evaluation demonstrates that \ToolName\ achieves high precision and recall with significant gains in efficiency, particularly when leveraging embedding models for semantic interpretation. Furthermore, the results highlight the practical impact of our approach in enhancing library-aware pointer analysis and taint analysis. By bridging the gap between informal documentation and formal analysis, \ToolName\ provides a robust and scalable solution for understanding library semantics. }

\bibliographystyle{ACM-Reference-Format}
\balance
\bibliography{sample-base}

\appendix

\end{document}


	\title[\ToolName]{\ToolName: Inferring API Aliasing Specifications from Library Documentation via Neurosymbolic Optimization}

	\ccsdesc[500]{Theory of computation~Program verification}
	\ccsdesc[300]{Software and its engineering~Software defect analysis}
	\ccsdesc[500]{Social and professional topics~Software maintenance}
	\ccsdesc[500]{Software and its engineering~Software maintenance tools}
	
	
	\maketitle
	
	\section{Temperature Sensitivity}
	To quantify the temperature sensitivity,
	we evaluate \ToolName\ under different temperature settings, including 0.1, 0.4, 0.7, 1.0, and 1.3.
	In the first group, we set the temperature of first stage $t_1$ to 0.7 and conduct the second stage of prompting under the five temperature settings.
	In the second group, we fix the temperature of second stage $t_2$ as 0.7 and set $t_1$ to the five different temperatures.
	
\begin{figure*}[h]
		\centering
		\includegraphics[width=\linewidth]{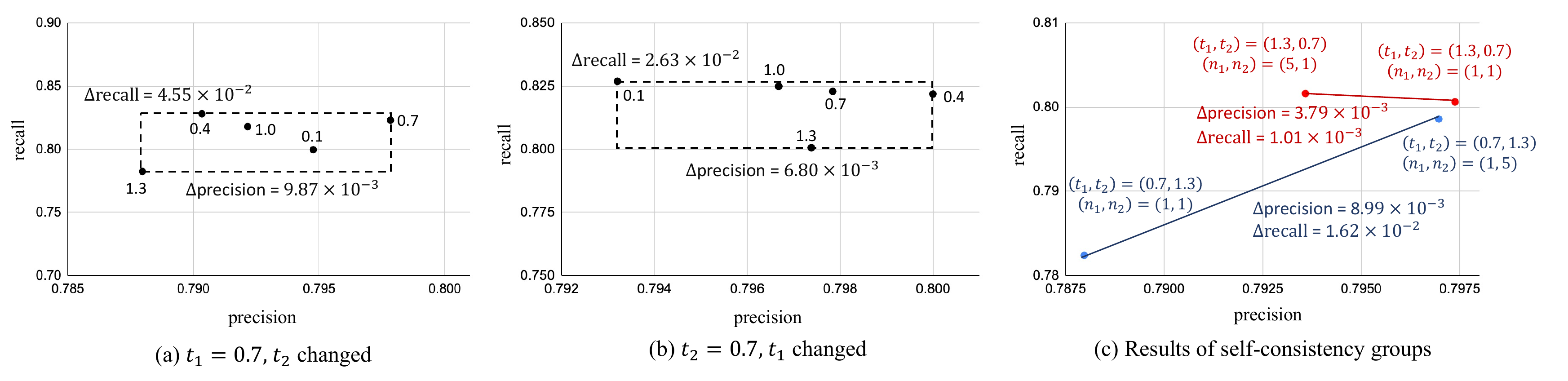}
		\caption{The results of \ToolName\ under different temperature settings}
		\label{fig:temperature}
	\end{figure*}

	Figure~\ref{fig:temperature}(a) and (b) show the precision and recall of \ToolName\ under different temperature settings \wu{what is temperature setting?}.
	Notably, the precision and recall can decrease slightly when the temperatures increase in the two stages,
	which is consistent with our intuition.
	The ranges of the precision and recall are only $9.87 \times 10^{-3}$/$6.80 \times 10^{-3}$ and $4.55 \times 10^{-2}$/$2.63 \times 10^{-2}$ in the first/second group, respectively.
	Furthermore, we also adopt the self-consistency strategy in the two stages respectively~\cite{DBLP:conf/iclr/0002WSLCNCZ23}.
	Considering that the experiments would introduce a large token cost,
	we repeat the LLM queries five times under two temperature settings $(0.7, 1.3)$ and $(1.3, 0.7)$,
	and use the maximum voting to choose the answers.
	Here $(t_1, t_2)$ indicates that the temperatures of the two stages are $t_1$ and $t_2$, respectively.
	Our experimental results show that the precision and recall are 79.70\% and 79.86\%, respectively, under the temperature setting $(0.7, 1.3)$.
	Similarly, \ToolName\ obtains 79.36\% precision and 80.16\% recall under the temperature setting $(1.3, 0.7)$.
	Hence, the results under the two settings are also very closed to the ones without adopting self-consistency strategy, 
	which demonstrates that our approach is not sensitive to the temperature setting.
	
	\bibliographystyle{ACM-Reference-Format}
	\balance
	\bibliography{sample-base}
	
	\appendix